\documentclass[sigplan,noacm,10pt]{acmart}
\renewcommand\footnotetextcopyrightpermission[1]{}
\usepackage{tikz}
\usepackage{amsmath}

\usepackage[framemethod=TikZ]{mdframed}
\usepackage{listings}
\usepackage{tcolorbox}

\usepackage{subcaption}
\usepackage{multirow}
\usepackage{xspace}
\usepackage{enumitem}
\usepackage{xargs}
\usepackage{soul}
\usepackage{colortbl}
\usepackage{bm}

\usepackage{cleveref}

\usepackage{pifont}

\usepackage{colortbl}

\newcommand{\cmark}{{\color{green!60!black}\ding{51}}}

\newcommand{\pname}{\textit{$\varphi$-Bench}\xspace}

\newcommand{\observationbox}[4]{%
  \vspace{#1}%
  \begingroup 
  \setlength{\fboxsep}{3.5pt}
  \noindent\makebox[\linewidth][c]{
    \fbox{%
      \parbox{#2}{%
        \linespread{0.95}\selectfont 
        #3\par 
      }%
    }%
  }%
  \endgroup
  \vspace{#4}%
}

\newcommand{\observation}[3][O]{%
  \observationbox{0.5ex}{0.96\linewidth}{\textls[0]{{\bf Takeaway (#1{#2}):}~{#3}}}{1ex}%
}

\definecolor{edge1color}{RGB}{54, 57, 61}
\definecolor{edge2color}{RGB}{108, 142, 191}
\definecolor{edge3color}{RGB}{184, 84, 80}
\definecolor{face1color}{RGB}{251, 217, 103}
\definecolor{face2color}{RGB}{218, 232, 252}
\definecolor{face3color}{RGB}{248, 206, 204}

\newcommand*\circleb[4][black]{\tikz[baseline=(char.base)]{
    \node[shape=circle,draw=#1,inner sep=0.5pt,fill=#2,text=#3,scale=0.95] (char) {#4};}}

\setcopyright{none}

\renewcommand\footnotetextcopyrightpermission[1]{}

\begin{document}

\title{Benchmarking LLMs on File System Design\\ and Implementation}

\author{Yuqi Xue}
\authornote{Co-primary authors.}
\email{yuqixue2@illinois.edu}
\affiliation{%
  \institution{University of Illinois}
  \city{Urbana-Champaign}
  \country{USA}
}

\author{Daixuan Li}
\authornotemark[1]
\email{daixuan2@illinois.edu}
\affiliation{%
  \institution{University of Illinois}
  \city{Urbana-Champaign}
  \country{USA}
}

\author{Jian Huang}
\email{jianh@illinois.edu}
\affiliation{%
  \institution{University of Illinois}
  \city{Urbana-Champaign}
  \country{USA}
}

\begin{abstract}

Large Language Models (LLMs) are fundamentally transforming computer system research and development.
As we employ LLMs in file system (\textit{fs}) development, it is essential to understand their capabilities, limitations, and operational efficiency for domain-specific tasks. 
We present \pname{}, an LLM benchmarking framework for \textit{fs}-specific tasks.
To facilitate benchmarking, we develop six types of tasks in \pname{}: basic understanding, basic implementation, performance modeling, debugging, optimization, and new feature development.
Each type emphasizes different LLM capabilities: instruction following, knowledge recall, reasoning, or coding. To create high-quality tasks while achieving broad coverage with minimal human effort, we develop a new AI-assisted task generation pipeline in addition to expert-written and textbook-adapted tasks. With 505 tasks in \pname{}, we conduct an empirical study with both open source (DeepSeek-V4-Flash, GLM-5.1, and MiniMax-M2.7) and proprietary (Claude-Opus-4.7, GPT-5.2, and Gemini-3.1-Pro) LLMs. Our study discloses the model efficiency for different tasks, causes of failed \textit{fs} tasks, and techniques for mitigating LLM failures. 
We will open source \pname{} to facilitate public research on using LLMs for \textit{fs} development.

\end{abstract}

\settopmatter{printacmref=false,printccs=false,printfolios=true}
\maketitle
\pagestyle{plain}

\section{Introduction}
\label{sec:intro}


Recently, large language models (LLMs) have been increasingly employed for computer system research and development, including software engineering, hardware development (RTL coding), OS configuration tuning, kernel debugging, and OS verification \cite{jimenez2023swe,rtllm,autoos24,mathai2024kgym,osvbench}.
Similar to other domain-specific engineering tasks that benefit from LLMs, developing and maintaining a file system (\textit{fs}) is often labor-intensive and time-consuming, as modern file systems have evolved into large and complex codebases. 
Hence, it is desirable to ask the question: can we employ LLMs to help reduce the human effort in file system development?


While recent coding models and agents~\cite{jimenez2023swe,openai-codex,anthropic-claude-code,cursor2024} have shown great promise for general software engineering, they mainly target code completion tasks, and application-level development and debugging.
It is unclear how well they can handle \textit{fs} design and implementation tasks.
A recent work SYSSPEC~\cite{specfs:fast25} shows that LLMs can generate \textit{fs} code from human-written specifications, but the LLM still relies on precise specifications written by human developers. It remains unclear whether LLMs can support broader and more difficult \textit{fs} development tasks that require them to 
explore design trade-offs and make design choices.

To effectively use LLMs for \textit{fs} development, we must first understand their capabilities and limitations in this domain.
Although many LLM benchmarks have been developed to evaluate domain-specific engineering tasks, from mathematical reasoning and general software engineering to scientific research~\cite{hle-benchmark,hendrycks2021measuring,lu2024mathvista,chen2021codex,jimenez2023swe,prakash2025quarch,rtllm,thakur2023verigen,mathai2024kgym,osvbench}, they cannot be used to evaluate LLMs on \textit{fs} development tasks, due to the fundamental differences in the domain knowledge, task types, development goals, and evaluation methodologies. 


\noindent
\textul{\textit{The \mbox{\pname{}} framework.}}
We develop \pname{} (Phi-Bench)\footnote{In the name \pname{}, $\varphi$ is pronounced ``phi,'' echoing ``file'' in file systems.}, the first benchmarking framework for evaluating LLMs on \textit{fs} design and implementation.
\pname{} consists of 505 Q\&A and coding tasks (see examples in \Cref{fig:task_examples}). The tasks cover all major components and functions in a production-grade file system (\Cref{tab:question_distribution}) and are organized into six types: basic understanding, basic implementation, performance modeling, debugging, optimization, and new feature development (\S\ref{sec:taxonomy}).
They represent typical tasks in the \textit{fs} development lifecycle.
Each type examines different LLM capabilities.


To ensure both high quality and broad coverage with reduced human effort, we develop a new AI-assisted task generation pipeline, 
alongside 
human expert-written and textbook-adapted tasks (\S\ref{sec:ai-generation}).
The key principle is to let an LLM draft and review the tasks, while keeping human experts in the loop for reviewing tasks and writing comments to guide the LLM to revise tasks, 
and making final accept/reject decisions.

To minimize human review effort,
each task is generated with two steps (\Cref{fig:task_construction_design_flow}).
First, we ask the LLM to generate \textit{task outlines}, which serve as ``proposals'' 
for easier human review
and must be approved before generating the full tasks.
The generation is grounded by OS textbooks and Linux kernel source code 
to ensure the generation
is not biased 
toward the model's own knowledge.
Second, we ask the LLM to expand the approved outlines to full tasks, including the task description, reference answer, and test harness.
For coding tasks, the test harness provides a self-contained \textit{fs} environment that implements the APIs and data structures for the task.
It initializes and warms up the involved \textit{fs} data structures before evaluating the answer.
Each full task first goes through an autonomous self-revision loop, where the LLM reviews and revises the task 
itself 
under a set of rules.
Then, the task will be reviewed: it is either accepted, rejected, or commented by human experts for LLM revision.

All tasks in \pname{} 
undergo double-blind peer review by human experts
to validate task quality.
Our AI-generated tasks achieve review scores comparable to expert-written and textbook-adapted tasks, and human reviewers cannot reliably distinguish 
AI-generated 
from human-created tasks (\S\ref{sec:task_assessment}).
We will open source \pname{} and its construction pipeline,
enabling future studies to adapt the pipeline to build high-quality benchmarks for other domains.


\noindent
\textul{\textit{Empirical study with \mbox{\pname{}}.}}
We use \pname{} to evaluate six frontier LLMs, including both open-source (DeepSeek-V4-Flash~\cite{DeepSeekV4}, GLM-5.1~\cite{glm2025}, and {MiniMax-M2.7}~\cite{minimaxm27}) and proprietary models (Claude-Opus-4.7~\cite{anthropic2025claude46}, GPT-5.2~\cite{openai2025gpt52}, and Gemini-3.1-Pro~\cite{google2025gemini3pro}).
Our study covers model performance (pass rate), output stability, failure cause analysis, and token/API cost (\S\ref{sec:overall_perf}--\S\ref{sec:model_cost}).
We also evaluate
representative prompt and context engineering techniques, 
including RAG, self-review, and error feedback (\S\ref{sec:solvability}).
We summarize our key takeaways as follows.

\begin{itemize}[leftmargin=*, nosep]
\vspace{1ex}
\item Current frontier LLMs perform well on simple task types, with the best pass rates reaching 95.8\%, 87.4\%, and 88.6\% on basic understanding, basic implementation, and performance modeling tasks.
This is expected as textbook knowledge is covered in the pretraining corpus, and modern LLM training emphasizes coding and reasoning.
However, their performance drops sharply on debugging (61.6\%), performance optimization (37.6\%), and new feature development (41.9\%) tasks.
Even though a model possesses relevant domain knowledge and general coding and reasoning capabilities, it may not be able to correctly apply them to solve \textit{fs} design and implementation tasks (\S\ref{sec:overall_perf}).

\item The dominant failures are not simple instruction following mistakes or missing domain knowledge.
Instead, they are caused by (1) misuse of semantically similar but functionally irrelevant knowledge ({19.1}\% of all failures), (2) logical inference errors ({22.3}\%), (3) missing edge cases due to incomplete reasoning ({14.5}\%), and (4) functionally correct code with suboptimal performance ({17.5}\%) (\S\ref{sec:failure_analysis}).

\item For most models, token usage is dominated by LLM reasoning.
However, longer reasoning traces do not necessarily mean better performance (pass rate), as they often reflect redundant reconsideration of trivial design decisions rather than deeper analysis of the task.
The API cost is generally positively correlated with pass rate: low-cost models such as DeepSeek-V4-Flash can be 16$\times$--107$\times$ cheaper than the best-performing models across task types, despite the 3.5\%--27.1\% lower pass rates (\S\ref{sec:model_cost}).

\item Because LLM generation is stochastic, a model may solve a task in one trial but fail in another.
Resampling (running a task multiple times and picking the best result) improves pass rate by up to 21\%.
However, 75\%--79\% of failures remain unresolved, showing that most failures are persistent capability failures rather than unlucky samples (\S\ref{sec:solvability}).

\item To mitigate failures, the effective prompt/context engineering techniques include (1) prompting the model to self-review its recalled knowledge, reasoning trace, and solution; and (2) providing compiler/runtime errors and the number of passed/failed unit tests to the model for iterative refinement.
These techniques improve the pass rate
from 46\%/64\% to 83\%/89\% for DeepSeek-V4-Flash/Gemini-3.1-Pro.
The remaining failures are dominated by incomplete reasoning and suboptimal implementations (\S\ref{sec:solvability}).
\vspace{1ex}
\end{itemize}

We anticipate \pname{} to benefit the community. 
First, our study provides practical guidance for both using and improving LLMs in \textit{fs} development.
For \textit{fs} developers, our study provides insights for building an agentic workflow with separate agents for gathering knowledge, generating design, writing code, and reviewing.
For model builders, our study suggests two optimization goals for future LLMs: systematic reasoning about edge cases and performance-aware code generation.
Second, beyond evaluating current models, \pname{} can serve as a dataset for fine-tuning \textit{fs}-specific LLMs.
Third, our AI-assisted task construction pipeline can be adapted to other domains to reduce the human effort in building domain-specific benchmarks and datasets (\S\ref{sec:discussion}).
Overall, we make the following contributions in this paper. 

\begin{itemize}[leftmargin=*, nosep]
\vspace{1ex}
    \item We construct \pname{}, 
    the first 
    LLM benchmarking framework on
    \textit{fs} design and implementation, with a 
    fine-grained taxonomy of \textit{fs} development tasks and LLM capabilities.
    
    \item We design an AI-assisted task generation pipeline 
    to produce high-quality \textit{fs} benchmark tasks with reduced manual drafting effort. Our methodology would inspire the benchmark construction for other domains.  
    
    \item We conduct a thorough empirical study with \pname{} with popular frontier LLMs. We expect our study results would shed light on the LLM for \textit{fs} research and development. 
    
\end{itemize}





\section{Background and Motivation}
\label{sec:bkg}

\subsection{Large Language Models}
\label{sec:bkg:llm}





Large language models (LLMs) are auto-regressive generative models that predict the next word (or sub-word unit, called a \emph{token}) given all preceding tokens~\cite{brown2020language,kaplan2020scaling}.
To solve downstream tasks beyond next-token prediction, modern LLMs are trained and evaluated for four basic capabilities:

\noindent
\textul{\textit{Instruction following:}} The LLM needs to understand the task written in natural language and follow the instructions (e.g., design constraints and output format) in the prompt~\cite{wei2021finetuned,ouyang2022training}.

\noindent
\textul{\textit{Knowledge recall:}} The LLM needs to retrieve correct and relevant facts encoded in its parameters (model weights), such as domain-specific concepts, thus, it can rely on the recalled knowledge to solve the given task~\cite{petroni2019language,jiang2020knowlanguagemodelsknow,kadavath2022language}.

\noindent
\textul{\textit{Reasoning:}} The LLM can perform chain-of-thought (CoT) reasoning, which generates intermediate reasoning tokens to decompose a complex problem into sub-tasks and solve each sub-task before producing the final answer~\cite{wei2022chain,kojima2023largelanguagemodelszeroshot}.

\noindent
\textul{\textit{Coding:}} 
Given a detailed design specification or a design derived from its own reasoning, 
the LLM can generate correct and efficient code for the intended functions~\cite{chen2021codex,li2023starcoder,roziere2023code}.

\subsection{Using LLMs for File System Development}







The long history of file system (\textit{fs}) development has produced large and complex codebases~\cite{ffs:tocs84,rosenblum1992lfs,xfs:usenix96,ceph:osdi06,lee2015f2fs,btrfs:tos13}.
Taking Linux as an example, \textit{fs} components span across multiple layers of the OS kernel and involve numerous critical functionalities, from the virtual file system (VFS) layer, page cache, core \textit{fs} logic (e.g., metadata and data indexing), to the block layer and device drivers (see our taxonomy in \Cref{tab:question_distribution}).
As file systems continue to evolve, developing and maintaining them becomes increasingly difficult and costly, requiring intensive human labor and time for adding new features, optimizing performance, and patching bugs~\cite{fsstudy:fast13,crashmonkey:osdi18,kim2019hydra,bhat2017scaling}.

Although coding models and agents~\cite{jimenez2023swe,openai-codex,anthropic-claude-code,cursor2024} have shown great promise for general software engineering tasks, they mainly target code completion and application-level development.
It remains unclear how well they can handle \textit{fs} design and implementation.
SYSSPEC~\cite{specfs:fast25} takes a closer step by using human-written specifications to guide LLMs in generating \textit{fs} code, however, writing such specifications 
requires substantial domain expertise and human effort, and the model primarily translates a well-specified design into code rather than making design 
decisions.
As a result, how to apply LLMs to 
\textit{fs} development remains an open 
problem.


\begin{figure*}[t]
    \includegraphics[width=\linewidth]{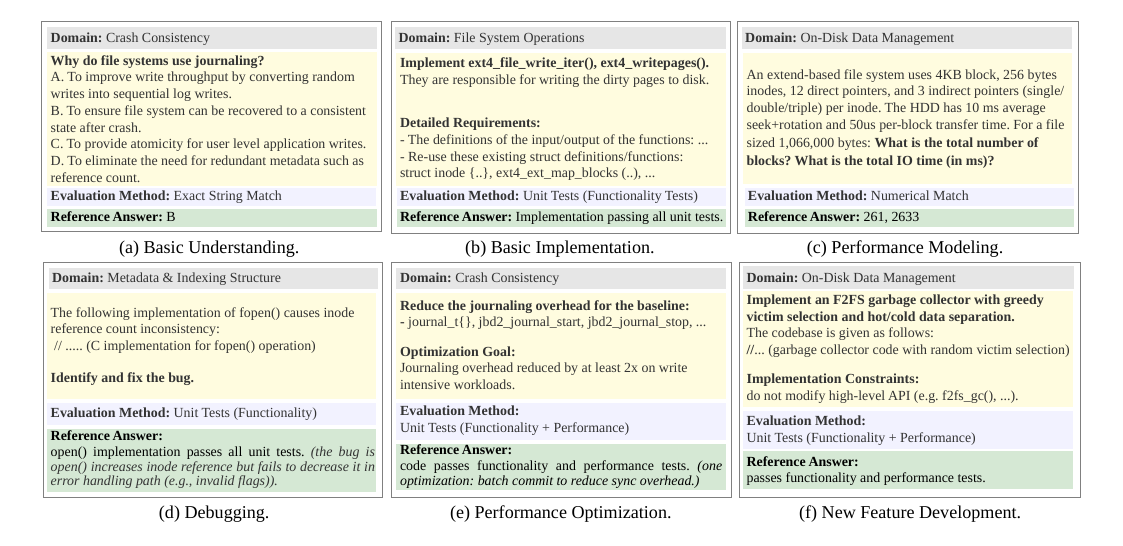}
    \caption{Representative tasks in \pname{}.}
    \label{fig:task_examples}
\end{figure*}

\subsection{\textls[-6]{Benchmarking LLMs on File System Development}}

To use LLMs for \textit{fs} development, we need to 
understand their capabilities and limitations in this domain.
Recent domain-specific LLM benchmarks cover domains 
such as mathematical reasoning~\cite{hle-benchmark,hendrycks2021measuring,lu2024mathvista}, general software engineering~\cite{chen2021codex,jimenez2023swe}, computer architecture~\cite{prakash2025quarch}, hardware (RTL) coding~\cite{rtllm,thakur2023verigen}, and OS debugging and verification~\cite{mathai2024kgym,osvbench}.
However, they cannot be directly used to evaluate LLMs for \textit{fs} development, due to the fundamental differences in the required domain knowledge, task types, development goals, and evaluation methodologies. Moreover, most benchmarks focus on gathering a collection of domain-specific tasks and mainly aim to provide an overall accuracy metric for the model under evaluation, rather than a fine-grained breakdown of the model's capabilities and limitations.

In this work, we aim to build a new benchmarking framework to systematically quantify an LLM's capabilities and limitations on \textit{fs} design and implementation. 
We hope that our study will inspire future research and development of using LLMs for \textit{fs} development, such as training specialized models and developing customized LLM agents.







\section{\pname{} Construction Methodology}

We build \pname{}, the first benchmark to evaluate LLM capabilities in \textit{fs} understanding, design, and implementation.
We first present a taxonomy to characterize the task types in the \textit{fs} development lifecycle and the model capabilities each task type requires
(\S\ref{sec:taxonomy}).
Based on our taxonomy, we create textbook-adapted, expert-written, and AI-generated tasks.
We build an AI-assisted task generation pipeline grounded by OS textbooks and Linux kernel code and supervised by human experts, which scales the benchmark construction while guaranteeing task quality (\S\ref{sec:ai-generation}).
All tasks are peer-reviewed by human experts.
Our AI-generated tasks achieve quality comparable to human-created tasks (\S\ref{sec:task_assessment}). 


\begin{table}[t]
    \centering
    \small
    \caption{Model capabilities tested by each task type in \pname{}. ``\cmark'' denotes that a capability is tested by a task type. }
    \label{tab:capability-mapping}
    \setlength{\tabcolsep}{4pt}

    {\footnotesize
    \begin{tabular}{@{}l@{ }l l@{ }l@{}}
    I: & Instruction Following & K: & Knowledge Recall \\
    R: & Reasoning             & C: & Coding \\
    \end{tabular}\par}
    \vspace{0.5ex}
    
    \begin{tabular}{@{} l c c c c @{}}
    \toprule
    \footnotesize \textbf{Task Type} & 
    \makebox[1em][l]{\rotatebox[origin=bl]{0}{\footnotesize\begin{tabular}[b]{@{}l@{}}\textbf{I}\end{tabular}}} & 
    \makebox[1em][l]{\rotatebox[origin=bl]{0}{\footnotesize\begin{tabular}[b]{@{}l@{}}\textbf{K}\end{tabular}}} & 
    \makebox[1em][l]{\rotatebox[origin=bl]{0}{\footnotesize\textbf{R}}} & 
    \makebox[1em][l]{\rotatebox[origin=bl]{0}{\footnotesize\textbf{C}}} \\
    \midrule
    Basic Understanding (BU)      & \cmark & \cmark &  --    &  --    \\
    Basic Implementation (BI)     & \cmark & \cmark &  --    & \cmark \\
    Performance Modeling (PM)     & \cmark & \cmark & \cmark &  --    \\
    Debugging (DE)                & \cmark & \cmark & \cmark & \cmark \\
    Performance Optimization (PO) & \cmark & \cmark & \cmark & \cmark \\
    New Feature Development (NF)  & \cmark & \cmark & \cmark & \cmark \\
    \bottomrule
    \end{tabular}
\end{table}

\subsection{\pname{} Taxonomy} 
\label{sec:taxonomy}





To analyze LLM capabilities and limitations in \textit{fs} development, 
\pname{} organizes tasks along two axes: 
which stage in the \textit{fs} development lifecycle it represents and what model capabilities it tests.
As shown in \Cref{tab:capability-mapping}, the lifecycle axis (rows) yields six task types that cover different development activities, such as debugging, developing new features, and optimizing performance.
The capability axis (columns) shows whether each type assesses instruction following, knowledge recall, reasoning, coding, or a combination of capabilities.

\noindent
\textbf{Basic Understanding (BU)} tasks (\Cref{fig:task_examples}a) test whether a model has the basic \textit{fs} knowledge (i.e., knowledge recall) needed before performing complex design and implementation tasks. They use multiple-choice, true/false, or short calculation formats, allowing the benchmark to isolate knowledge recall from reasoning and coding capabilities.

\begin{table}[t]
  \centering
  \small
  \setlength{\tabcolsep}{4pt}
  \setlength{\aboverulesep}{0.4ex}
  \setlength{\belowrulesep}{0.4ex}
  \caption{Distribution of tasks by type, domain, and source.}
  \label{tab:question_distribution}
  \footnotesize
  \begin{tabular}{llrr}
  \toprule
   & & \textbf{Count} & \textbf{\%} \\
  \midrule
  \multirow{6}{*}{\textbf{Type}} & Basic Understanding & 62 & 12.3\% \\
   & Basic Implementation & 57 & 11.3\% \\
   & Performance Modeling & 74 & 14.7\% \\
   & Debugging & 85 & 16.8\% \\
   & Performance Optimization & 124 & 24.6\% \\
   & New Feature Development & 103 & 20.4\% \\
  \midrule
  \multirow{15}{*}{\textbf{Domain}} & Virtual File System & 50 & 9.9\% \\[0pt]
   & \quad {\footnotesize \textit{File System API}} & {\footnotesize 50} & {\footnotesize 9.9\%} \\
   & Memory Management & 43 & 8.5\% \\[0pt]
   & \quad {\footnotesize \textit{Page Cache}} & {\footnotesize 43} & {\footnotesize 8.5\%} \\
   & File System Core & 264 & 52.3\% \\[0pt]
   & \quad {\footnotesize \textit{File System Operations}} & {\footnotesize 35} & {\footnotesize 6.9\%} \\[-1pt]
   & \quad {\footnotesize \textit{Metadata \& Indexing}} & {\footnotesize 62} & {\footnotesize 12.3\%} \\[-1pt]
   & \quad {\footnotesize \textit{On-Disk Data Layout}} & {\footnotesize 36} & {\footnotesize 7.1\%} \\[-1pt]
   & \quad {\footnotesize \textit{Concurrency}} & {\footnotesize 44} & {\footnotesize 8.7\%} \\[-1pt]
   & \quad {\footnotesize \textit{Crash Consistency}} & {\footnotesize 37} & {\footnotesize 7.3\%} \\[-1pt]
   & \quad {\footnotesize \textit{Data Integrity}} & {\footnotesize 50} & {\footnotesize 9.9\%} \\
   & Block Layer & 86 & 17.0\% \\[0pt]
   & \quad {\footnotesize \textit{I/O Scheduling}} & {\footnotesize 46} & {\footnotesize 9.1\%} \\[-1pt]
   & \quad {\footnotesize \textit{Virtual Block Devices}} & {\footnotesize 40} & {\footnotesize 7.9\%} \\
   & Device Driver & 62 & 12.3\% \\
  \midrule
  \multirow{4}{*}{\textbf{Source}} & AI-Generated & 305 & 60.4\% \\
   & Human-Created & 200 & 39.6\% \\[-1pt]
   & \quad {\footnotesize \textit{Expert-Written}} & {\footnotesize 93} & {\footnotesize 18.4\%} \\[-1pt]
   & \quad {\footnotesize \textit{Textbook-Adapted}} & {\footnotesize 107} & {\footnotesize 21.2\%} \\
  \midrule
  \multicolumn{2}{l}{\textbf{Total}} & \textbf{505} & \textbf{100\%} \\
  \bottomrule
  \end{tabular}
\end{table}

\noindent
\textbf{Basic Implementation (BI)} tasks (\Cref{fig:task_examples}b) test 
an LLM's ability to translate a well-specified design into C code.
The prompt provides step-by-step workflow/pseudocode, function interfaces, available APIs, and implementation constraints.
The model must write code and pass unit tests.
This 
separates implementation (i.e., coding) skill from design reasoning.
\pname{}'s BI tasks differ from generic coding tasks because the model needs
basic \textit{fs} concepts to generate the correct implementation, 
such as how each field in an inode should be queried or modified to implement the provided design.

\noindent
\textbf{Performance Modeling (PM)} tasks (\Cref{fig:task_examples}c) assess a model's quantitative reasoning capability about \textit{fs} performance.
Tasks describe a workload pattern, hardware configuration, or code snippet, and the model must analyze performance, compare designs, and make design decisions. 

\noindent
\textbf{Debugging (DE)} tasks (\Cref{fig:task_examples}d) test the model's ability to identify and fix bugs.
The model is given a buggy code snippet and a bug report that includes the inputs that trigger the bug and the expected vs. observed behavior.
The model must understand the bug report, reason about the root cause of the bug, and implement the correct patch.


\noindent
\textbf{Performance Optimization (PO)} tasks (\Cref{fig:task_examples}e) require the model to improve an existing implementation to satisfy a performance target while preserving correctness. Unlike PM tasks, which focus on reasoning about the optimization strategy, PO tasks also require the model to implement the optimization. Our tasks cover diverse metrics, such as latency, throughput, I/O amplification, and disk-space utilization.

\noindent
\textbf{New Feature Development (NF)} tasks (\Cref{fig:task_examples}f) ask the model to add new functionality to an existing file system, such as a new POSIX API or data management policy. Unlike BI tasks, NF tasks state the new feature's goal rather than a step-by-step implementation plan. 
For performance related features (e.g., adding hot/cold data separation mechanism), unit tests include correctness and performance tests.
NF tasks involve understanding the required feature, reasoning about functionality and performance, and generating correct code.

Overall, our taxonomy enables an ablation study of different capabilities of LLMs. BU, BI, and PM isolate knowledge recall, reasoning, and coding, respectively. DE, PO, and NF test the combination of these capabilities in various end-to-end development tasks. Instruction following is shared by all categories because each task requires parsing the task specification and producing the specified output format.

\pname{} also organizes tasks by domain (\Cref{tab:question_distribution}). These domains cover major layers in the \textit{fs} stack. We also include relevant OS layers: page cache, block layer, and device drivers, as developing a file system often requires interacting with them. 
The domain/task-type distribution roughly reflects the 
topic coverage
in OS textbooks and Linux source code, as they grounded our benchmark construction 
(\S\ref{sec:ai-generation}).

\subsection{Benchmark Construction}
\label{sec:ai-generation}

\begin{figure*}[t]
    \centering
    \includegraphics[width=\linewidth]{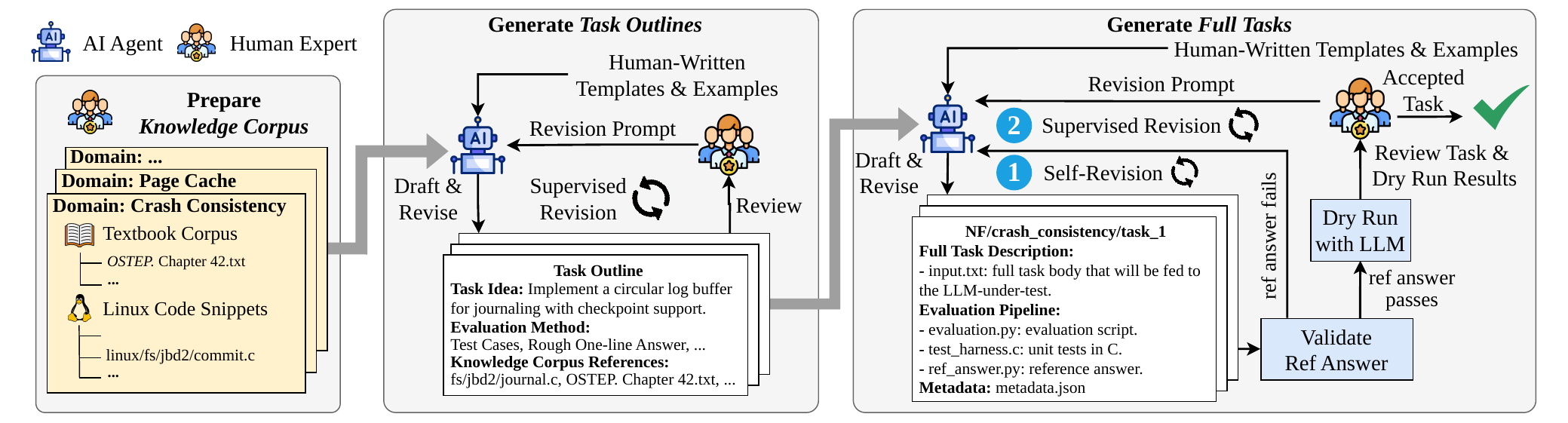}
    \caption{AI-assisted task generation pipeline in \pname{}.}
    \label{fig:task_construction_design_flow}
\end{figure*}

\pname{} needs to achieve two goals: high task quality and broad coverage of domains and task types.
A standard way to construct an LLM evaluation benchmark is to crowd-source from a large number of human authors~\cite{prakash2025quarch,hle-benchmark}.
However, while human experts can write high-quality tasks, this method is both time-consuming and labor-intensive.
Another way is to adapt from textbook exercises and exams.
However, these exercises are often too simple and not designed for testing coding capability.
To achieve both high quality and broad coverage while minimizing human effort, \pname{} introduces a new AI-assisted task generation pipeline in addition to the expert-written and textbook-adapted tasks.
\Cref{tab:question_distribution} reports the number of tasks from each source.

\vspace{-0.55em}
\subsubsection{Textbook-Adapted Tasks}
We use LLM to extract all \textit{fs}-related exercises from two popular OS textbooks~\cite{ArpaciDusseau23-Book,silberschat8operating}.
We include only multiple-choice, true/false, calculation, and coding exercises.
For Python coding exercises, we prompt the LLM to convert them into Linux kernel-style C code.
We omit open-ended free-response questions because 
deterministic test harnesses cannot automatically evaluate them\footnote{Though it is possible to use LLM to evaluate free-response questions (i.e., LLM-as-a-judge), this introduces biased or incorrect evaluation results~\cite{szymanski2024limitations}.}.

\vspace{-0.55em}
\subsubsection{Expert-Written Tasks}
Five graduate students who have experience in systems research (i.e., ``human experts'') manually created the 93 tasks in \pname{}, taking $\approx${580} person-hours in total.
All tasks are peer-reviewed and revised in a double-blinded manner to cross-validate task quality.

The human-based workflow has five steps: 
(1) Pick a \textit{fs} component/functionality based on the author's own knowledge or related material such as OS textbooks, real \textit{fs} code, or research papers.
(2) Formulate a task outline, including the core idea, evaluation method, and expected answer.
(3) Based on the outline, draft the task, reference solution, and test harness (for coding tasks).
(4) Verify the reference solution passes the tests and dry-run the task with LLMs to find potential issues, such as ambiguity, leaked answer cues, or buggy test harness.
(5) Revise the task to fix all issues.

\vspace{-0.55em}
\subsubsection{AI-Generated Tasks}
Our AI-assisted task generation pipeline is inspired by the above human workflow.
The key principle is to let an LLM agent perform the labor-intensive drafting/revision while keeping humans in the loop for reviewing tasks, writing comments to guide AI to revise the tasks, and making final decisions to accept/reject a task.
\Cref{fig:task_construction_design_flow} shows the pipeline, which consists of three stages.

\noindent
\textul{\textit{(1) Prepare knowledge corpus.}}
We first construct the \textit{fs} knowledge corpus, which will be fed to the LLM agent in 
later stages.
It consists of textbook chapters and sections~\cite{ArpaciDusseau23-Book,silberschat8operating}, which provide conceptual knowledge and design principles, and code snippets from Linux kernel source code~\cite{linux-kernel-code}, which provide real-world coding examples. We curate a knowledge corpus for each domain in \Cref{tab:question_distribution}, and each is curated manually for factual correctness and relevance.

This reduces hallucination and minimizes bias by providing external factual knowledge. Without it, 
the LLM would generate tasks only from its own knowledge; consequently, the resulting task will only cover what the model already knows and fail to evaluate the model objectively.


\noindent
\textul{\textit{(2) Generate task outlines.}}
For each target domain and task type, the LLM agent generates a set of candidate \textit{task outlines}.
These outlines serve as ``proposals'' that must be approved by human supervisors before being expanded to full tasks.
Each task outline contains the core task idea, the intended evaluation method, the high-level reference answer, and the knowledge corpus entries used to derive this task.
An outline is much more concise than a full task and hence requires minimal human effort to review (typically $\leq$200 words need to be reviewed for each outline).
Generating an outline before the full task is critical for improving final task quality and reducing human efforts, as we can quickly identify ``promising'' outlines that are more likely to yield high-quality tasks.

We prompt the LLM agent to generate at least {20} outlines for each domain and task type.
The generation is grounded by the corresponding knowledge corpus, the task outline template,
and 3--5 example human-written outlines.
Each outline is reviewed by a human expert for correctness, relevance, and duplicates.
An outline is either accepted/rejected, or the reviewer will write a revision prompt to guide the AI to revise it.
This \textit{supervised revision} process is iterative, and it finishes when all outlines are either accepted or rejected.

\noindent
\textul{\textit{(3) Generate full tasks.}}
For each accepted task outline, the LLM agent expands the task idea into a \textit{full task}, which includes multiple files: the full task description that will be fed to the LLM-under-test (\texttt{input.txt}), the unit tests written in C code if this is a coding task (\texttt{test\_harness.c}), the reference answer (\texttt{ref\_answer.py}), and the evaluation script that takes the model output, invokes the unit tests, and gives the final score (\texttt{evaluation.py}).

For each coding task, the test harness provides a self-contained \textit{fs} environment that implements the APIs and data structures required by the task. It warms up all involved \textit{fs} state, such as in-memory caches and on-disk structures, before running the unit tests. The harness code is grounded by real Linux kernel code in the knowledge corpus and validated by human experts during task review.

For each generated full task, the LLM agent first launches an autonomous \textit{self-revision} loop (\circleb[cyan]{cyan}{white}{1}) to examine the task with a set of rules: the task description matches the evaluation script, the evaluation script and the reference answer must compile and run, the reference answer passes all unit tests, and the task description does not leak answer cues.
The agent will revise a task iteratively to address these issues.

After self-revision, a task undergoes \textit{supervised revision} (\circleb[cyan]{cyan}{white}{2}). We dry-run the task with an LLM, after which human supervisors review both the task and the dry-run results.
The dry run ensures task failures reflect incorrect answers rather than task flaws, such as ambiguous instructions, buggy evaluation scripts, or faulty unit tests.
\Cref{tab:eval-rubric} shows the review rubric. Note that technical depth is not considered for task acceptance; it is collected to validate that \pname{} covers different difficulty levels. 

\begin{table}[t]
    \centering
    \small
    \setlength{\tabcolsep}{6pt}
    \setlength{\aboverulesep}{0.2ex}
    \setlength{\belowrulesep}{0.4ex}
    \caption{Review rubrics used by human experts in \pname{}'s benchmark construction and validation.}
    \label{tab:eval-rubric}
    \footnotesize
    \begin{tabular}{@{}ll@{}}
        \toprule
        {\footnotesize\textbf{Technical Soundness}} & {\footnotesize\textbf{Clarity}} \\
        \midrule
        \textbf{1:} Complete hallucination & \textbf{1:} Unintelligible \\
        \textbf{2:} Contains factual errors & \textbf{2:} Idea present, logic unclear \\
        \textbf{3:} No issue & \textbf{3:} Mostly clear \\
        & \textbf{4:} Very clear \\
        \midrule
        {\footnotesize\textbf{Technical Depth}} & {\footnotesize\textbf{Overall Quality}} \\
        \midrule
        \textbf{1:} Trivial basic recall & \textbf{1:} Reject \\
        \textbf{2:} Shallow familiarity & \textbf{2:} Major revision \\
        \textbf{3:} Moderate knowledge & \textbf{3:} Minor revision \\
        \textbf{4:} Multi-step reasoning & \textbf{4:} Accept as is \\
        \textbf{5:} Cross-layer reasoning & \textbf{5:} Best task candidate \\
        \bottomrule
    \end{tabular}
\end{table}

\subsection{Benchmark Quality Assessment}
\label{sec:task_assessment}

\begin{figure}[t]
    \centering
    \includegraphics[width=\linewidth]{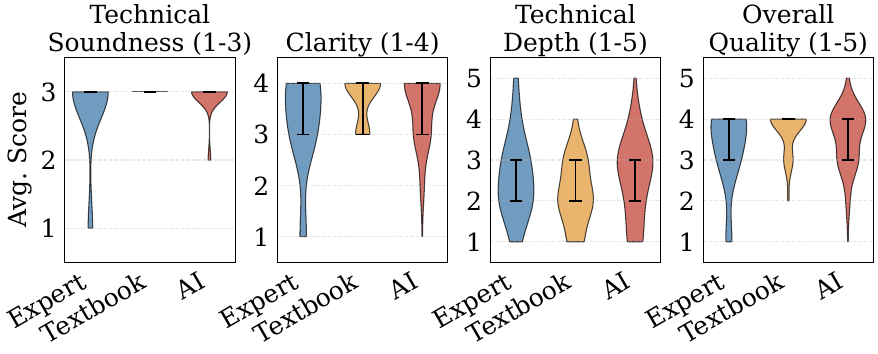}
    \caption{Review scores of tasks in \mbox{\pname{}}. Vertical lines show the 25th/75th percentiles. Textbook-adapted tasks always score 3 in technical soundness, as they are adapted from publication-quality exercises.}
    \label{fig:human_vs_ai}
\end{figure}



\begin{table}[t]
    \centering
    \small
    \setlength{\tabcolsep}{5pt}
    \renewcommand{\arraystretch}{0.9}
    \setlength{\aboverulesep}{0.2ex}
    \setlength{\belowrulesep}{0.4ex}
    \caption{Double-blind test results for distinguishing between human-written and AI-generated tasks. Human testees are asked to annotate the source of 90 sampled tasks.}
    \label{tab:blind-test-results}
    \begin{subtable}[t]{0.4\linewidth}
      \centering
      \caption{Annotation accuracy.}
      \label{tab:blind-test-accuracy}
      \footnotesize
      \begin{tabular}{lc}
          \toprule
          {\footnotesize\textbf{Source}} & {\footnotesize\textbf{Accuracy}} \\
          \midrule
          Human & 50.5\,\% \\
          AI & 48.9\,\% \\
          \midrule
          {\footnotesize\textbf{Overall}} & {\textbf{49.5\,\%}} \\
          \bottomrule
      \end{tabular}
    \end{subtable}\hfill
    \begin{subtable}[t]{0.56\linewidth}
      \centering
      \caption{Confusion matrix.}
      \label{tab:blind-test-confusion}
      \footnotesize
      \begin{tabular}{l cc}
          \toprule
          {\footnotesize\textbf{Source}} & \multicolumn{2}{c}{\footnotesize\textbf{Human Annotation}} \\
          \cmidrule(lr){2-3}
           & Human & AI \\
          \midrule
          Human & \textbf{20.0\,\%} & 19.6\,\% \\
          AI    & 30.9\,\%          & \textbf{29.5\,\%} \\
          \bottomrule
      \end{tabular}
    \end{subtable}
\end{table}

To coordinate the review process, we built a website to collect expert-written tasks/reviews and version-control tasks across review/revision cycles.
All tasks 
are reviewed by $\geq$3 experts in a double-blinded manner.
Final inclusion requires unanimous acceptance from all reviewers, and any revised task must undergo another review cycle.
We will publish the website to crowdsource more tasks and public reviews.

\Cref{fig:human_vs_ai} compares the review scores of tasks across all three sources before any revision (all non-rejected tasks are revised and ultimately accepted).
Every domain in \pname{} contains at least three non-AI-generated tasks for a fair comparison.
Since our pipeline filters out poor task outlines early, AI-generated tasks in the first revision iteration already achieve quality comparable to human-created ones.
While expert-written and AI-generated tasks span a broad range of technical depth, textbook-adapted ones are simpler, as they focus on basic concepts for educational purposes rather than full-fledged design \& implementation problems.

To further validate \pname{}'s quality, we conducted a double-blind experiment asking four graduate students working on systems research to classify 30 randomly selected tasks per source (90 sampled tasks in total) as either human- or AI-generated.
\Cref{tab:blind-test-results} shows that testers correctly identified the source with roughly 50\% accuracy --- equivalent to random guessing.
This confirms that our AI-generated tasks are indistinguishable from those created by human experts.

\section{Empirical Study with \pname{}}
\label{sec:empirical_study}

In this section, we use \pname{} to evaluate six frontier LLMs.
Our results show that current LLMs still struggle with the most complex design and implementation tasks, and that prompt and context engineering methods improve but do not close this gap.
We highlight our key takeaways in boxes.

\subsection{Experimental Setup}

We evaluate both open-source (DeepSeek-V4-Flash~\cite{DeepSeekV4}, GLM-5.1~\cite{glm2025}, and {MiniMax-M2.7}~\cite{minimaxm27}) and proprietary models (Claude-Opus-4.7~\cite{anthropic2025claude46}, GPT-5.2~\cite{openai2025gpt52}, and Gemini-3.1-Pro~\cite{google2025gemini3pro}).
For GLM-5.1 and MiniMax-M2.7, we use Together AI API~\cite{togetherai2024}.
For other models, we use their official APIs.
All models use default temperatures and maximum reasoning efforts.
The experiments run on a server with Intel Xeon E5-2687W v4 (24 cores) and 96 GB memory.
To account for sampling variance of LLMs~\cite{kadavath2022calibration}, we run five independent trials per task and report their average results, unless otherwise specified.
To account for the task count difference across task types (\Cref{tab:question_distribution}), the overall pass rate is computed as the average of the per-type pass/fail rate weighted by the task counts.

\begin{figure}[t]
    \centering
    \includegraphics[width=\linewidth]{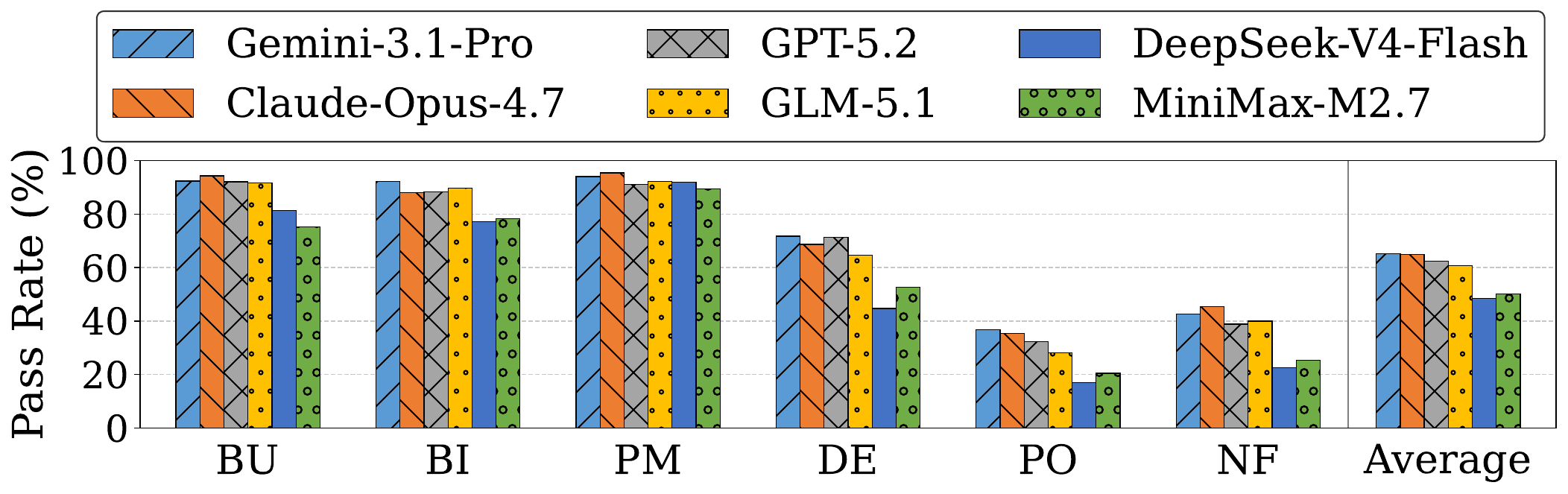}
    \caption{Pass rate of models across task types.}
    \label{fig:question_distribution_avg_qtype}
\end{figure}

\begin{table*}[t]
    \centering
    \setlength{\tabcolsep}{3pt}
    \caption{Definition of failure types across all failed trials for all models.}
    \label{tab:failure-taxonomy}
    \footnotesize
    \begin{tabular}{@{}
      >{\raggedright\arraybackslash}p{0.03\textwidth}
      >{\raggedright\arraybackslash}p{0.08\textwidth}
      >{\raggedright\arraybackslash}p{0.16\textwidth}
      >{\raggedright\arraybackslash}p{0.68\textwidth}
    @{}}
        \toprule
        Code & Capability & Failure Type & Definition \\
        \midrule
        F1 & Instruction Following & Specification Violation &
        The prompt states an explicit constraint, such as an output format, a fixed function signature, or a specific API usage, but the model's response violates it and cannot be parsed, compiled, or evaluated. \\\hline
        
        F2 & Knowledge Recall & Irrelevant/Misleading Knowledge &
        The model recalls a piece of knowledge that is factually correct but cannot be directly applied to solve the task, and this recalled fact misleads the reasoning process, leading to a wrong final solution. \\\hline
        
        F3 & Knowledge Recall & Incorrect or Missing Knowledge &
        The model hallucinates or misses \textit{fs} facts, making it impossible to reason about the correct solution based on these facts. \\\hline
        
        F4 & Reasoning & Logical Inference Error &
        The model starts from a correct and sufficient premise, but makes a logical inference error, such as an incorrect causal relation, leading to a wrong final answer. \\\hline
        
        F5 & Reasoning & Incomplete Reasoning &
        The model starts from a correct and sufficient premise, but misses some necessary reasoning steps. As a result, the final solution misses critical edge cases, such as a boundary check or error handling path. \\\hline
        
        F6 & Coding & Reasoning-Solution Mismatch &
        The reasoning process concludes with a valid solution, but the submitted answer does not follow the proposed solution, such as using a different algorithm or data structure. \\\hline
        
        F7 & Coding & Suboptimal Implementation &
        The functional design and optimization strategies are correct, but implementation introduces unnecessary overhead, such as redundant \texttt{malloc()/free()} and suboptimal choice between \texttt{qsort()} vs. insertion sort. \\\hline
        
        F8 & Coding & Generic Coding Bugs &
        The code compiles, but fails at runtime due to generic, non-\textit{fs}-specific coding bugs, such as a segmentation fault, double free, use-after-free, or heap corruption. \\
        
        \bottomrule
    \end{tabular}
\end{table*}

\subsection{Overall Model Performance}
\label{sec:overall_perf}

\Cref{fig:question_distribution_avg_qtype} shows the average performance (pass rate) on different task types.
All models perform well on BU (up to 95.8\%), BI (up to 87.4\%), and PM (up to 88.6\%) tasks. 
This suggests that frontier models have strong textbook-level \textit{fs} knowledge, can implement simple and well-specified \textit{fs} functionality, and can analytically model \textit{fs} performance when the task is abstracted out from a complex codebase.
The results are expected: OS textbooks are well-covered in pretraining corpora (BU), modern frontier models are explicitly optimized for coding (BI), and they can apply basic quantitative reasoning to known \textit{fs} concepts (PM).

\observation[T]{1}{
LLMs perform well on knowledge recall (BU), simple coding (BI), and performance modeling (PM). This is consistent with how they are trained: pretraining covers OS textbook knowledge, and modern training pipelines emphasize coding and reasoning.
}

All models perform substantially worse on DE, PO, and NF tasks, achieving up to {61.6\% on DE, 37.6\% on PO, and 41.9\% on NF}.
These tasks require the model to jointly recall \textit{fs} knowledge, reason about functionality and performance, and implement correct code. The difficulty is pronounced for PO and NF, where the model must preserve correctness while changing the design to satisfy performance and functionality requirements. We analyze the failed cases in detail in \S\ref{sec:failure_analysis}.


\observation[T]{2}{
LLMs struggle with complex file system 
tasks that require multiple capabilities. 
Possessing domain knowledge, reasoning, and coding capability does not guarantee a model can apply them correctly to solve concrete design and implementation problems.
}

\begin{figure}[t]
    \centering
    \includegraphics[width=\linewidth]{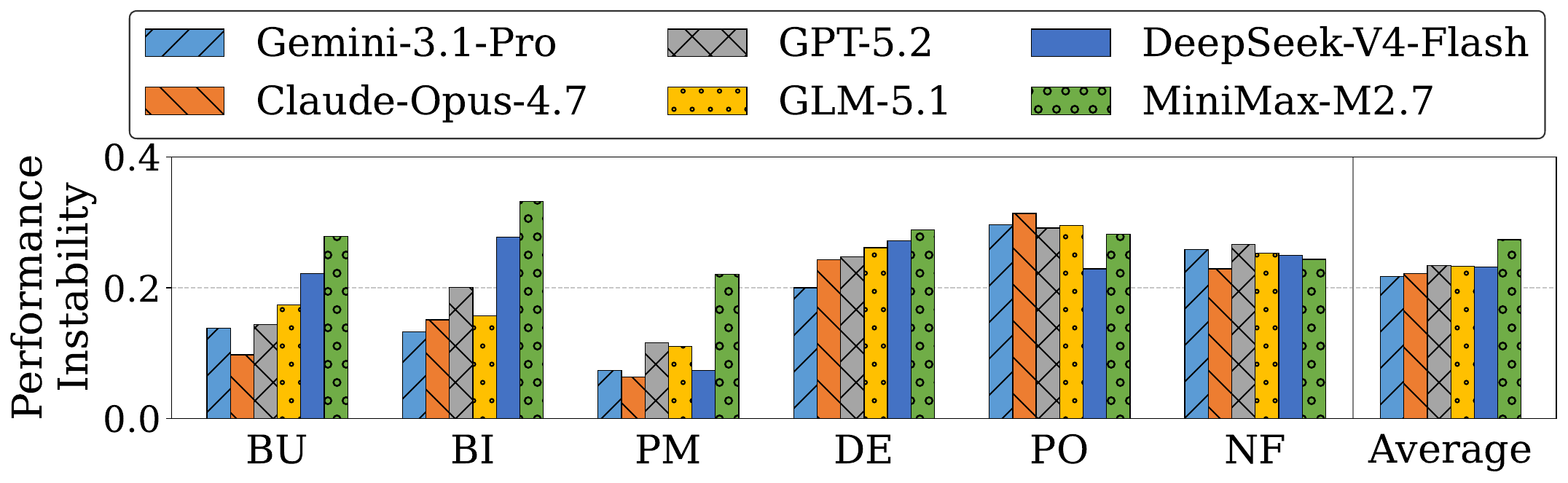}
    \caption{Performance instability of models across task types. Instability is measured as the pooled standard deviation~\cite{montgomery2017design} of 5 repeated trials, with pass scored as 1.0 and fail as 0.0. Lower instability values indicate the model consistently performs well or poorly on the target task type.}
    \label{fig:question_distribution_pooled_std}
\end{figure}

LLMs are inherently stochastic: the same input prompt can produce different outputs across runs, so a model may answer a task correctly in one trial but fail in another.
We quantify the impact of output instability of LLMs in Figure~\ref{fig:question_distribution_pooled_std}.

A model that achieves a higher pass rate also answers tasks more consistently.
This is because when an LLM knows the correct answer, it tends to produce next-token distributions heavily skewed toward the correct answer, so the token corresponding to the correct answer has substantially higher probability to be selected than tokens corresponding to incorrect answers~\cite{kadavath2022calibration}.
Intuitively, the model is more ``confident'' when it knows the answer.
When the LLM is uncertain about the answer, it produces a distribution that is more uniform across correct and incorrect tokens, so it ``guesses'' a random answer.
This trend is obvious for BU, BI, PM, and DE tasks.
For the most difficult PO and NF tasks, all models suffer from low pass rates and have relatively high instability.

\observation[T]{3}{
Models with higher pass rates produce more stable and consistent outputs across multiple trials due to the stochastic sampling nature of LLMs.
}

\begin{figure}[t]
    \centering
    \includegraphics[width=\linewidth]{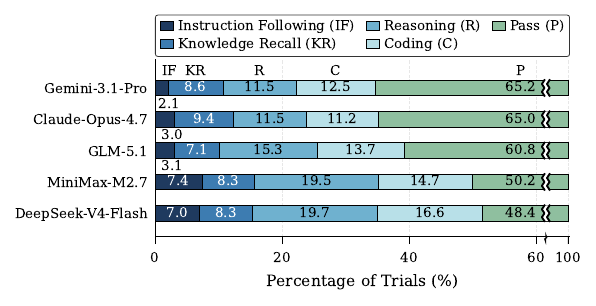}
    \caption{The percentage of tasks failed due to different failure reasons listed in \Cref{tab:failure-taxonomy}. We exclude GPT-5.2, as its reasoning process is not available. 
    }
    \label{fig:fail_rate_stacked_bar}
\end{figure}

\begin{figure*}[t]
    \centering
    \vspace{-2ex}
    \includegraphics[width=\linewidth]{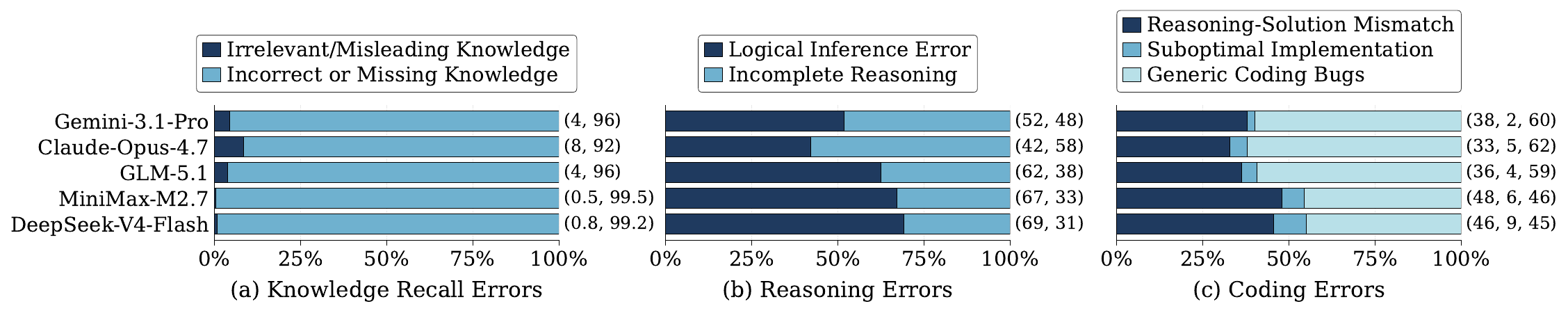}
    \caption{Failure type breakdown based on the taxonomy in \Cref{tab:failure-taxonomy}. We exclude GPT-5.2, as its reasoning process is unavailable.
    }
    \label{fig:error_breakdown}
\end{figure*}

\subsection{Failure Analysis}
\label{sec:failure_analysis}

In this section, we analyze how and why a model fails to correctly solve a task.
We manually inspected every failed trial and categorized their failures into 8 types (see \Cref{tab:failure-taxonomy}).
We omit GPT-5.2 as its API does not expose its reasoning process.
\Cref{fig:fail_rate_stacked_bar} shows the failure type breakdown.

\noindent\textbf{Instruction following (F1).}
Instruction following failures are the least common 
failure type (2.1\%--7.4\% of all trials). Most of them are due to model output instability, as resampling resolved 72\%--98\% of them
(\S\ref{sec:solvability}).
The rest 
reveal certain model-specific biases: for example, Claude-Opus-4.7 
includes non-existent headers
in a few tasks, and DeepSeek-V4-Flash sometimes includes Markdown syntax in the generated code, causing compilation failures.
These failures can be solved by providing 
error messages for iterative refinement (see \S\ref{sec:solvability}).

\observation[T]{4}{
Instruction following failures are the least common failure type in all failure types (2.1\%-7.4\%). These failures are due to output instability or model-specific biases. They can be resolved by retrying or feeding error messages back to the model for revision.
}



\noindent\textbf{Knowledge recall (F2-F3).}
Knowledge recall errors account for 7.1\%--9.4\% of all failures. They fall into two categories.

As shown in \Cref{fig:error_breakdown}a, the majority (92\%--99\%) of knowledge recall failures are due to the model recalling irrelevant or misleading facts (F2).
Even if the model has the correct knowledge, it may not be able to precisely recall it when needed. Instead, it may recall semantically similar knowledge, such as a memorized algorithm or 
code pattern,
and apply it directly without reasoning.
For example, in a BU task on deadlock detection for file locks, Gemini-3.1-Pro correctly recalls Linux's efficient chain-based deadlock detection mechanism in \texttt{fs/locks.c}.
However, in an NF task requiring the same deadlock detection within a file lock manager, the model recalls 
a costlier dependency graph DFS algorithm, failing the performance test.
To alleviate this, we can ask the model to carefully verify recalled knowledge before reasoning about the answer (see \S\ref{sec:solvability}).

\observation[T]{5}{
The majority of knowledge recall failures (92\%-99\%) are due to recalling semantically similar but unsuitable knowledge, which misleads the reasoning process and leads to an incorrect solution.
}

In \Cref{fig:error_breakdown}a, 0.5\%--8\% of knowledge recall failures are due to incorrect or missing knowledge (F3).
The missing knowledge is about specific production-level \textit{fs} implementations rather than textbook-level generic design principles.
For example, MiniMax-M2.7 knows that a write() to a file opened with O\_DIRECT (i.e., direct I/O) bypasses the page cache, but it does not recall that in ext4, the write also invalidates stale page cache data.
19\% of missing knowledge failures are resolved by resampling, indicating that they are due to output instability; the remaining ones can be addressed by providing the missing knowledge (e.g., via RAG) to the model to ground its output, or asking the model to verify its recalled knowledge and recall again when necessary (see \S\ref{sec:solvability}).

\observation[T]{6}{
The missing knowledge in frontier models is production-level file system implementation facts that a human developer acquires in real development experience, rather than textbook-level design principles.
}

\noindent\textbf{Reasoning (F4-F5).}
Reasoning failures mainly fall into two types: either the model makes a wrong logical deduction in its reasoning process given a correct knowledge context (F4), or its reasoning is incomplete, which typically leads to missing edge cases in
the final solution (F5). 
As shown in \Cref{fig:error_breakdown}b, both are significant within reasoning failures.

Logical inference error (F4) accounts for 42\%--69\% of all reasoning failures, as shown in \Cref{fig:error_breakdown}b.
For example, in an NF task that requires implementing XFS-style inode allocation, GLM-5.1 first proposes a valid design that manages inodes in groups, each group reserving a fixed-size inode-number space.
However, it then rejects this valid design, deeming the reserved space too large without calculating it.

We observe that some logical errors may be fixed later as the model performs iterative refinement of the solution in its reasoning process. In fact, across all trials where the reasoning process involves logical inference errors, {55}\% of them ultimately reach a correct final solution, despite significant token consumption (see \S\ref{sec:model_cost}). To reduce logical inference errors, prompting the model to do more self-refinement is a feasible way to improve reasoning quality (see \S\ref{sec:solvability}).

\observation[T]{7}{
42\%--69\% of reasoning failures are due to logical inference errors. However, in 55\% of the reasoning traces containing these errors, the model successfully self-corrects by iteratively verifying and refining its solution.
}

31\%--58\% of reasoning failures are due to incomplete reasoning (F5), which typically leads to missing edge cases in the final solution, even though the given context and recalled knowledge are sufficient for the model to derive all edge cases.
{For example, in a task on implementing a POSIX read(), GLM-5.1 does not reason about the boundary case where the read offset is greater than the file size.}


{This reveals a fundamental limitation of chain-of-thought (CoT) reasoning: CoT can help decompose a complex task into smaller, more solvable steps, but it does not guarantee that the model will systematically cover all edge cases}~\cite{brown2020language,wei2022chain,brown2025large}.
 Prompting the model to explicitly examine edge cases and verify the proposed solution can help alleviate this type of failure (see \S\ref{sec:solvability}).
To systematically eliminate this failure type in practice, we should leverage mature debugging, testing, and formal verification tools for file systems~\cite{kim2019hydra, mohan2018crashmonkey,  sigurbjarnarson2016yggdrasil}.

\observation[T]{8}{
31\%--58\% of reasoning failures are due to incomplete reasoning, which leads to missing edge cases in the generated code.  This is an intrinsic limitation of CoT reasoning: CoT can decompose a complex task into smaller, more solvable steps, but does not guarantee coverage. 
}

\begin{figure}[t]
\centering
\includegraphics[width=0.9\linewidth]{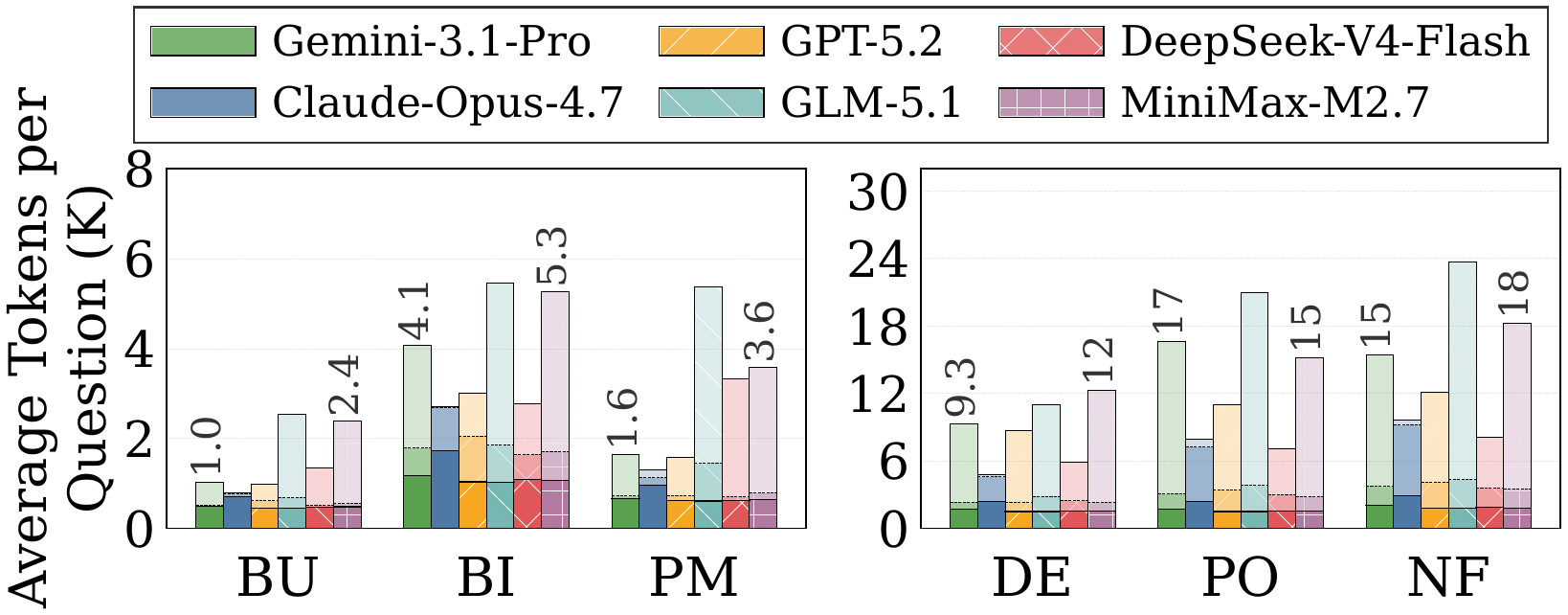}
\caption{Token usage per task by model and category (stacked bars: input at bottom, response in the middle, and thinking at the top). We mark the absolute values for Gemini-3.1-Pro and MiniMax-M2.7 for reference. 
}
\label{fig:token-usage-model}
\end{figure}

%




\noindent\textbf{Coding (F6-F8).}
Coding failures happen when a model recalls sufficient knowledge and performs correct reasoning to derive a valid design, but fails to implement it correctly. We categorize these into three types:

First, 33\%--48\% of coding failures are due to a mismatch between a correctly-reasoned solution and its implementation (F6), as shown in \Cref{fig:error_breakdown}c.
For example, in an NF task involving a write buffer queue, Claude-Opus-4.7 correctly reasons that the write counter should be incremented 
only for new queue entries, not for overwritten ones. 
However, the generated code still increments the counter on overwrite.

Second, 45\%--62\% of coding failures are due to suboptimal implementations (F7). For NF and PO tasks, which have performance requirements, the model may output a functionally and algorithmically correct solution,
but fail to implement it efficiently, such as sorting an already-sorted array or calling memcpy() twice on unchanged data.

Both F6 and F7 failures result from the stochastic nature of LLMs, which prevents any guarantee of perfect code generation.
However, these failures can often be mitigated using repeated sampling and self-code review (see \S\ref{sec:solvability}). 

Finally, 2\%--9\% of coding failures are generic, non-\textit{fs}-specific errors (F8), such as null/invalid pointer dereferences, double frees, use-after-free bugs, uninitialized variables, and off-by-one errors.
Since these often lead to compilation or runtime errors, a straightforward fix is to provide error messages to the LLM for refinement (see \S\ref{sec:solvability}).

\observation[T]{9}{
Even if a model derives a correct solution in its reasoning process, it may fail to 
generate correct and efficient code for the solution, 
due to the intrinsic probabilistic nature of LLMs. Such coding failures can often be addressed by repeated sampling, self-code review, or providing compilation or runtime error feedback.
}

\subsection{Cost Analysis}
\label{sec:model_cost}

In this section, we analyze the token usage and API cost (i.e., monetary cost) across models and task types. 



\noindent
\textbf{Token usage.}
As shown in Figure~\ref{fig:token-usage-model}, reasoning tokens account for 79.8--91.4\% of output tokens in all models, except for Claude-Opus-4.7 (only 9.5\%). 
The reasoning traces of open-source models\footnote{Because proprietary models such as Claude-Opus-4.7 and Gemini-3.1-Pro encrypt their original reasoning traces and expose only summaries, we focus on open-source models for the reasoning token usage analysis.} reveal substantial redundancy.
Even when a model solves a task correctly, it often overthinks the problem by repeatedly revisiting decisions after reaching a valid solution. For example, MiniMax-M2.7 spends over 47K tokens in a single trial, repeatedly reconsidering trivial factors like JSON-format constraints, whereas the core design decision only consumes {4.2K} reasoning tokens.
In contrast, Claude-Opus-4.7 only spends 251 tokens on the same task to derive the correct solution, indicating that the ``quality'' of reasoning matters more than the length.

\observation[T]{10}{
For most models, the token consumption is dominated by reasoning tokens (79.8--91.4\%).
Open-source models overthink by repeatedly revisit trivial design decisions, leading to high token usage.
As a result, more reasoning tokens do not necessarily lead to a higher pass rate; rather, the ``quality'' of reasoning matters more.
}


\begin{figure}[t]
    \centering
    \includegraphics[width=\linewidth]{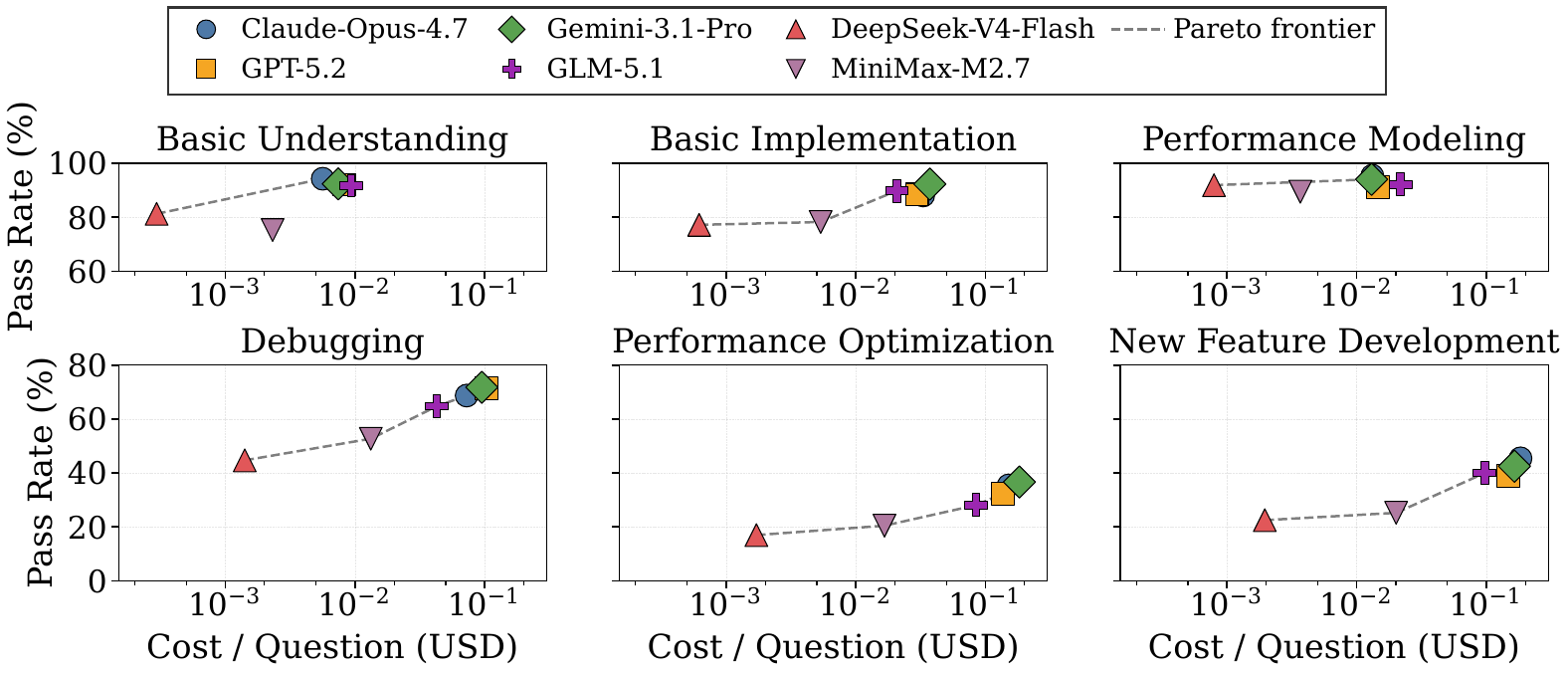}
    \caption{API cost vs. pass rate per task type. }
    \label{fig:cost-per-question}
\end{figure}

\noindent
\textbf{API (monetary) cost.}
Figure~\ref{fig:cost-per-question} shows the Pareto frontier of pass rate vs. cost for each task type\footnote{We use the following API prices per 1M input/output: \$5/\$25 (Claude-Opus-4.7), \$2/\$12 (Gemini-3.1-Pro), \$1.75/\$14 (GPT-5.2), \$0.98/\$3.08 (GLM-5.1), \$0.30/\$1.20 (MiniMax-M2.7), \$0.14/\$0.28 (DeepSeek-V4-Flash).}.
On each task type, models' end-to-end API costs are positively correlated with their performance (i.e., pass rates).
For example, DeepSeek-V4-Flash is {16$\times$--107$\times$} cheaper than the best-performing model across task types, despite its {3.5\%--27.1\%} lower pass rate.

\observation[T]{11}{
In general, the API cost of a model is positively correlated with its performance (i.e., pass rate).
}

\begin{figure}[t]
    \centering
    \includegraphics[width=\linewidth]{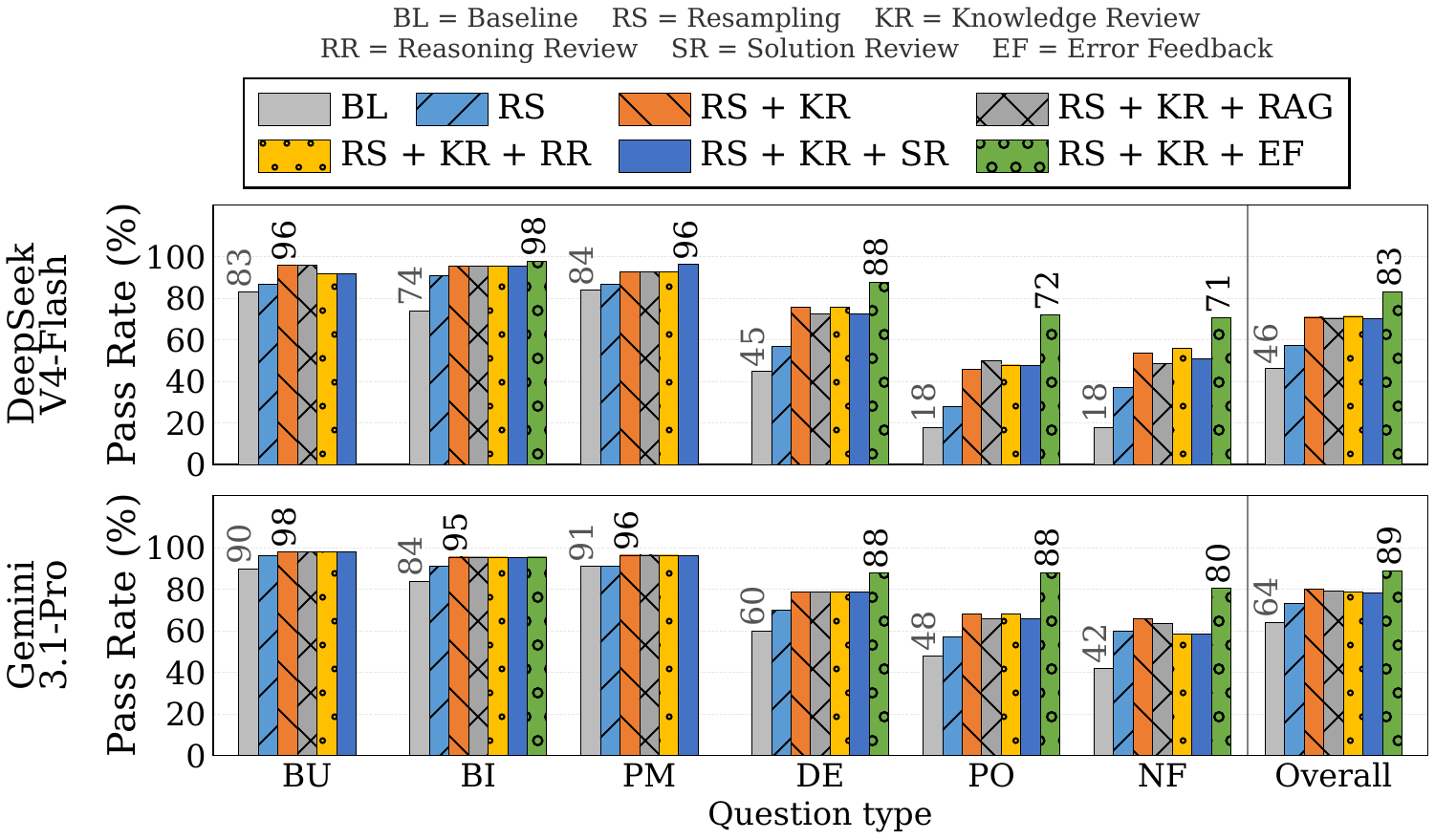}
    \caption{Pass rate improvement breakdown for each method in \S\ref{sec:solvability}. We show two models as examples. The observations also generalize to other models. In the legend, ``A + B'' means we apply methods A and B together.}
    \label{fig:context_engr_agg}
\end{figure}

\begin{figure}[t]
    \centering
    \includegraphics[width=\linewidth]{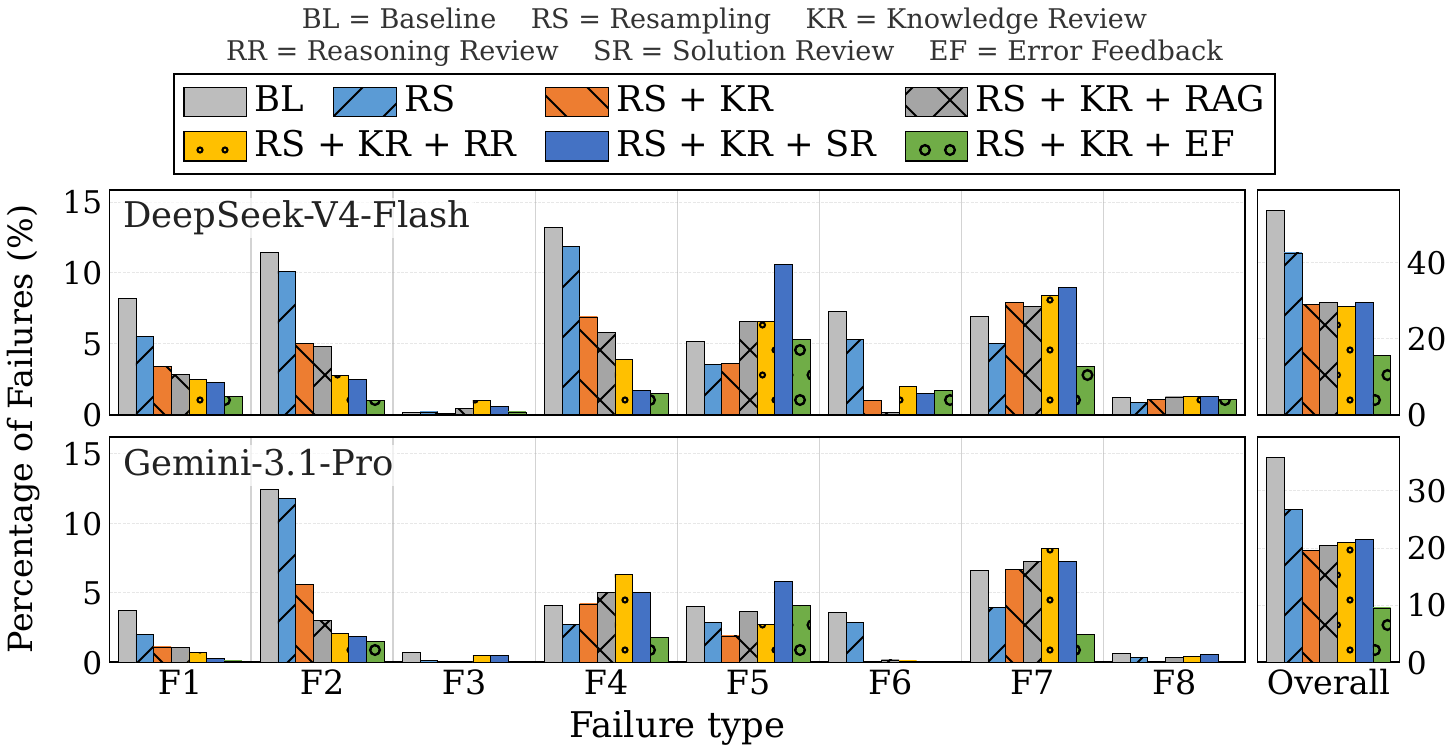}
    \caption{Percentage of failed trials after applying different combinations of prompt/context engineering methods.}
    \label{fig:context_engr_failure_type_breakdown}
\end{figure}

\begin{figure}[t]
    \centering
    \includegraphics[width=\linewidth]{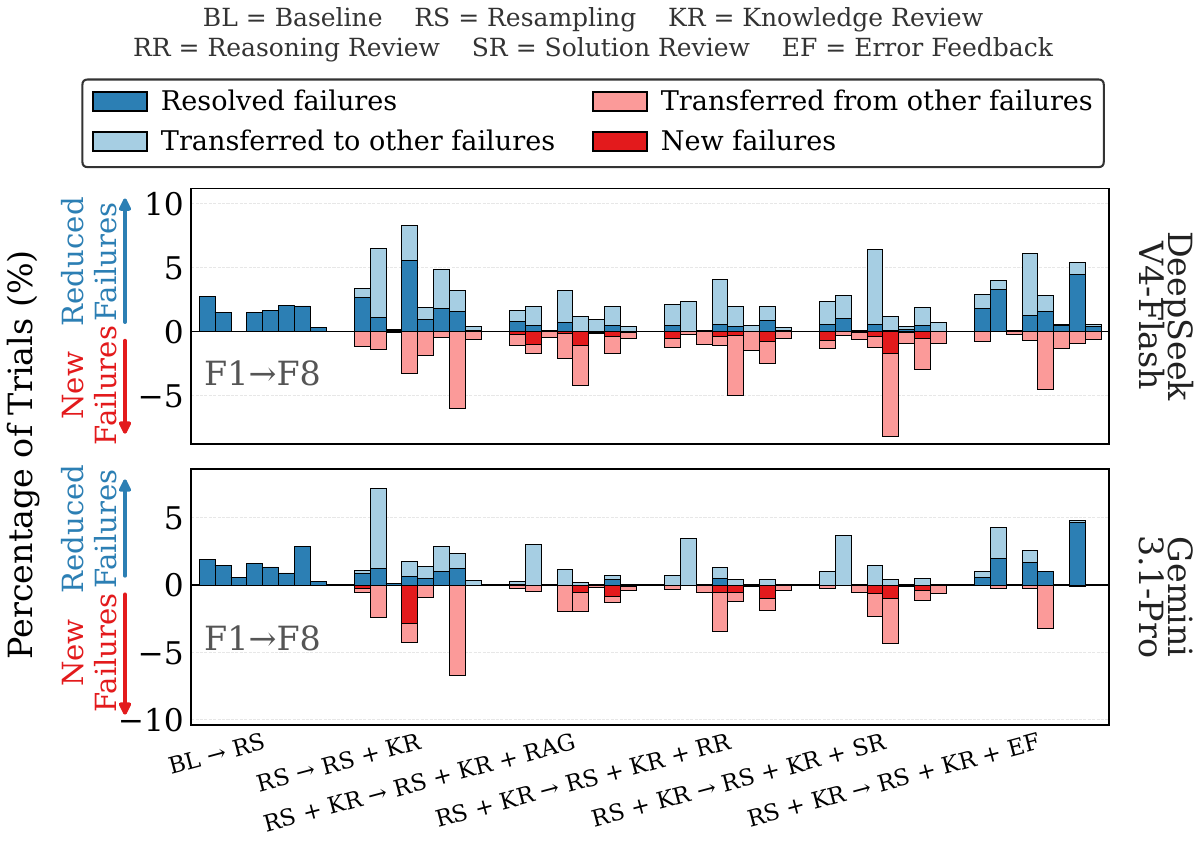}
    \caption{Failure type transitions after adding each method. For each transition \(A \to B\), the bars show the changes in F1--F8 failures (\Cref{tab:failure-taxonomy}). 
    Upward bars count failures in \(A\) that disappear in \(B\): the task either passes (``Resolved failures'') or fails with another failure type (``Transferred to other failures''). Downward bars count failures in \(B\) that were absent as in \(A\): they either come from another type (``Transferred from other failures'') or from a task that previously passed in \(A\) (``New failures''). Percentages are over all trials.}
    \label{fig:context_engr_failure_type_transition}
\end{figure}

\subsection{Mitigating Failures with Prompt Engineering and Context Engineering}
\label{sec:solvability}

To mitigate the failures discussed in \S\ref{sec:failure_analysis}, prompt engineering and context engineering~\cite{jiang2020knowlanguagemodelsknow,wei2022chain,brown2020language,kojima2023largelanguagemodelszeroshot,agarwal2024many} are popular approaches that improve LLM performance at inference time without updating model weights. In this section, we experiment with various methods that provide additional instructions or external knowledge to the model and evaluate their effectiveness in addressing each failure type:




\begin{itemize}[leftmargin=*, nosep]

\item \textbf{Resampling (RS)~\cite{wang2023selfconsistency}}: We repeatedly run each task for up to 12 times, stopping once the model produces a passing solution. Resampling beyond 12 trials does not help solve more tasks. This mitigates model output instability (see \Cref{fig:question_distribution_pooled_std}), targeting F1, F6, and F7 failures (see \S\ref{sec:failure_analysis}).

\item \textbf{Knowledge Self-Review (KR)}: 
We prompt the model to carefully verify that 
Factual knowledge in its reasoning process is correct and relevant to the task. This targets F2.

\item \textbf{Retrieval-Augmented Generation (RAG)}: We leverage RAG to provide extra knowledge to the model, targeting F3 failures. We build two RAG databases: (1) an OS textbook database~\cite{ArpaciDusseau23-Book,silberschat8operating}, and (2) an example code database that encodes \textit{fs}-related code from Linux~\cite{linux-kernel-code}. We retrieve the top-5 items from the two databases using hybrid BM25+semantic search~\cite{bm25,dpr,hybrid_retrieval,colbert} (a state-of-the-art RAG approach).
When applying KR+RAG together, we prompt the model to verify both the recalled and retrieved knowledge.

\item \textbf{Reasoning Self-Review (RR)}: We add instructions in the prompt to ask the model to review its reasoning process sentence by sentence to find and fix reasoning errors (F4 and F5 failures), and then regenerate the solution.

\item \textbf{Solution Self-Review (SR)}: We add instructions in the prompt to ask the model to review its solution (e.g., a code review) to fix F6--F8 failures before submitting the answer.

\item \textbf{Error Feedback (EF)}: We feed the compiler error message, runtime exception message, and the number of failed 
tests back to the model and ask it to refine its answer~\cite{chen2023selfdebug}, targeting F8 failures.
Similar to resampling, the error feedback loop runs for multiple iterations (up to {20} by default).

\end{itemize}
\vspace{0.4em}

\begin{figure}[t]
    \centering
    \includegraphics[width=\linewidth]{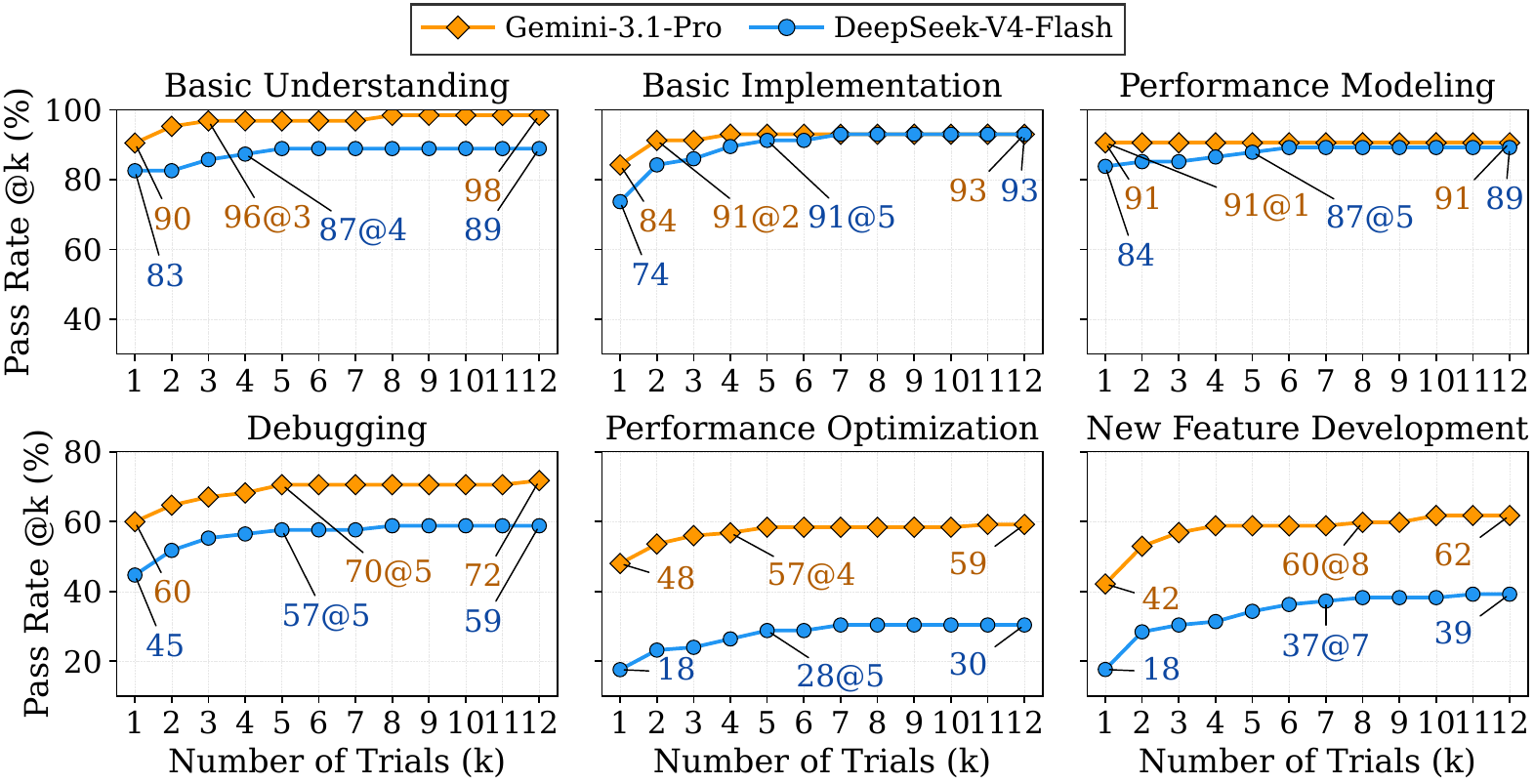}
    \caption{Overall pass rate at $k$ resampling runs (i.e., up to $k$ attempts for each task). We label the saturation point (``\{pass rate\}@\{$k$ runs\}'') at which more attempts improve the pass rate by $\leq$2\% (equivalent to 1--2 tasks per task type).
    }
    \label{fig:pass-rate-token-useage-multiple-run}
\end{figure}

Starting from the baseline, we progressively apply the above methods and examine whether each one improves the pass rate.
If a method improves the pass rate for any task type, we preserve it when adding the next method. If it does not affect (or even hurts) the pass rate, we discard it.

\Cref{fig:context_engr_agg} shows the overall pass rate improvement of different combinations of methods.
We focus on Gemini-3.1-Pro and DeepSeek-V4-Flash, as they represent the best-performing and the most cost-efficient models (see \S\ref{sec:model_cost}). The takeaways also generalize to other models.
Overall, resampling, knowledge self-review, and error feedback yield the most significant pass rate improvements; RAG and reasoning/solution self-review have no significant impact or even hurt the pass rate.
\Cref{fig:context_engr_failure_type_breakdown} shows the percentage of failures of each type when applying these methods, and \Cref{fig:context_engr_failure_type_transition} breaks down the changes in each failure type before and after adding a method.
We analyze each of them as follows.

\noindent\textbf{Resampling (RS).} Resampling improves the pass rate over the baseline ($k=1$) by up to 21\%.
As resampling mitigates model output instability, it helps reduce all types of failures (see \Cref{fig:context_engr_failure_type_breakdown}).
More unstable outputs result in larger gains; resampling is more effective for more difficult tasks (DE, PO, and NF) where models are more unstable (see \Cref{fig:question_distribution_pooled_std}).
Similarly, it is also more effective for models whose output is more unstable.
Intuitively, resampling allows the model to try different solutions, but it does not fundamentally improve a model's capability; {75}\%--{79}\% of failures remain unresolved.

As shown in \Cref{fig:pass-rate-token-useage-multiple-run}, the improvement of resampling diminishes as we increase the number of trials. For Gemini-3.1-Pro, pass rate is saturated at 1--5 trials for BU, BI, and PM tasks and 5--8 trials for DE, PO, and NF tasks.
In the remainder of this section, we configure the number of trials based on the saturation point for each task type in \Cref{fig:pass-rate-token-useage-multiple-run}.

\observation[T]{12}{
Resampling improves the pass rate by up to 21\% by mitigating output instability.
The improvement is greater for difficult tasks and for models that are more unstable.
However, substantial failures remain, as \textls[-3]{resampling cannot fundamentally improve model capability.}
}

\noindent\textbf{Knowledge Self-Review (KR).}
Prompting the model to verify all its recalled knowledge improves the overall pass rate by {13.5}\% for DeepSeek-V4-Flash and {6.9}\% for Gemini-3.1-Pro.
Knowledge recall failures (F2 and F3) are reduced by {50}\%--{53}\%, indicating that the model can self-correct some of the irrelevant or misleading knowledge during reasoning.
However, the effectiveness of such self-review is largely affected by model-specific biases.
For example, when prompted to self-review, Gemini-3.1-Pro over-critiques its recalled knowledge, falsely classifying correct knowledge as incorrect leading to reasoning failures (F4),
as shown in \Cref{fig:context_engr_failure_type_transition}.







\observation[T]{13}{
Prompting the model to self-review its recalled knowledge reduces the number of knowledge recall failures by {50}\%--{53}\%.
However, the effectiveness of self-review depends heavily on the model's reasoning capability and biases; the model may sometimes mistakenly classify correct knowledge as wrong.
}

As a side effect, knowledge self-review reduces other types of failures significantly.
The review instructions in the prompt implicitly encourage the model to review its entire reasoning process, double-check all recalled knowledge is relevant to the task specification, and perform more in-depth reasoning.
This eliminates up to {45}\% of instruction following failures, up to {32}\% of reasoning failures, and up to {11}\% of coding failures, although more suboptimal implementation failures (F7) become visible after fixing prior failures.

\observation[T]{14}{
As a side effect, the knowledge self-review instructions encourage the model to double-check the task specification, review the entire reasoning process and the solution, and perform more in-depth reasoning. This significantly reduces the number of instruction following, reasoning, and coding failures as well.
}

\noindent\textbf{RAG.}
RAG does not improve and even hurts the pass rate ({$-0.5\%$} for DeepSeek-V4-Flash and {$-1.0\%$} for Gemini-3.1-Pro).
This is because both models already have sufficient \textit{fs} knowledge and general coding capability (F3 failures are few).
Moreover, RAG suffers from intrinsic retrieval inaccuracies, which introduce more F2 failures (irrelevant/misleading knowledge).
Despite asking
to verify retrieved knowledge, 
models are overconfident about the correctness of RAG and directly apply textbook solutions without adapting them properly, leading to more missing edge cases (F5 failures) and suboptimal implementations (F7 failures), as shown in \Cref{fig:context_engr_failure_type_transition}.
Our observations align with recent studies, which suggest RAG may not always be the best way for retrieving external knowledge for coding tasks~\cite{pmlr-v235-wu24a, li2025impact, gu2025retrieve}.

\observation[T]{15}{
For frontier models, RAG does not offer significant improvements (and sometimes causes more failures due to retrieval inaccuracies); these models already have sufficient knowledge for most \textit{fs} development tasks.
}



\noindent\textbf{Reasoning Self-Review (RR) \& Solution Self-Review (SR).}
Neither reasoning self-review (RS+KR+RR) nor solution self-review (RS+KR+SR) improves the pass rate over knowledge self-review (RS+KR).
As a side effect of KR, the model already fixes many reasoning and coding failures (\textbf{T14}). 
As the model reviews its own reasoning and solution, its chain of thought becomes longer, and the model becomes more likely to make mistakes. 

\observation[T]{16}{
Since knowledge self-review already verifies the generated solution and intermediate reasoning steps, performing more self-review has diminishing or negative returns.
The model will likely make more mistakes as its chain of thought lengthens during further review.
}

\begin{figure}[t]
    \centering
    \includegraphics[width=\linewidth]{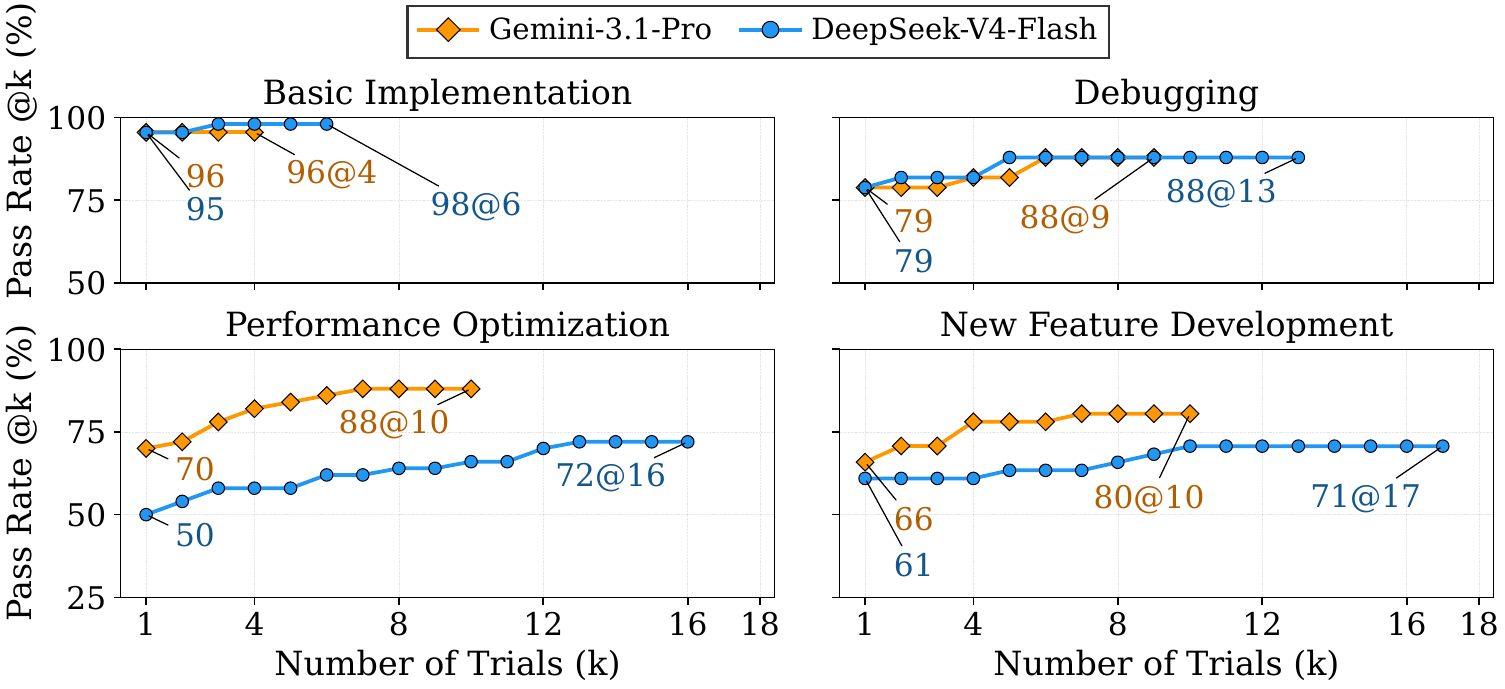}
    \caption{Pass rate at $k$ error feedback iterations. We only evaluate up to the saturation point (``\{pass rate\}@\{$k$ iterations\}'') at which 4 consecutive iterations do not further increase the pass rate by more than 1\%.
    }
    \label{fig:pass-rate-token-useage-error-feedback}
\end{figure}


\noindent\textbf{Error Feedback (EF)} improves the overall pass rate by {8.8}\%--{12.3}\%, with the largest gain on DE, PO, and NF tasks (+{9}\% to {26}\%).
Iterative refinement allows the model to try different solutions (similar to resampling), and feedback helps the model filter out wrong paths.
This helps resolve diverse types of failures, as shown in \Cref{fig:context_engr_failure_type_transition}.
However, the effectiveness of error feedback still relies on the model's ability to reason about feedback and propose solutions.
If a model cannot correctly interpret the feedback to propose a correct solution, more iterations yield diminishing returns (see \Cref{fig:pass-rate-token-useage-error-feedback}).
Hence, reasoning failures (F4 and F5) and suboptimal implementations (F7) dominate remaining failures.



\observation[T]{17}{
Providing the model with error feedback for iterative refinement improves the overall pass rate by up to {12.3}\%, as the feedback helps the model locate failures and try different solutions.
However, effectiveness is limited by the model's intrinsic reasoning capability,
leaving up to {58}\% of failures unresolved. 
}

\vspace{-1ex}
\section{Discussion}
\label{sec:discussion}

\noindent
\textbf{Implications for AI-driven \textit{fs} development.}
Our study provides takeaways for both \textit{fs} developers who use LLMs and ML researchers who develop future models.

For \textit{fs} developers, our study in \S\ref{sec:solvability} suggests a potential agentic workflow for \textit{fs} development, with separate agents for gathering and verifying \textit{fs} knowledge, deriving the design, writing the code, and conducting reviews.
The developers still need to provide comprehensive unit tests for error feedback and iterative refinement, and human supervision may be necessary when performance is critical or exhaustive edge-case coverage is required.
We wish to explore building agents for \textit{fs} design and implementation as future work.

For the ML community, the remaining gaps are systematic edge-case reasoning and performance-aware code generation.
Future models should therefore improve the ability to reason exhaustively about correctness constraints and translate designs into efficient implementations.
\noindent
\textbf{Using and extending \textit{$\bm{\varphi}$-Bench}.}
\pname{} provides representative \textit{fs} development tasks.
It can serve as a reference for evaluating new models and agents.
As each task in \pname{} includes a description, reference answer, and test harness, \pname{} can also serve as a dataset for post-training (supervised fine-tuning or reinforcement learning) LLMs.
To improve its coverage and quality, we will open source \pname{} and publish our website to collect more tasks and reviews.

\noindent
\textls[-3]{\textbf{Generalizability of \textit{$\bm{\varphi}$-Bench}'s construction pipeline.}}
Our benchmark construction pipeline can be
adapted to other domains.
Given the knowledge corpus, task templates, and review forms prepared by the domain experts, they can use a similar workflow to leverage AI agents to generate domain-specific tasks.
We hope our methodology inspires the development of more domain-specific benchmarks with 
reduced human effort.
We will open source our pipeline, including 
prompts, 
orchestration scripts, and website source code.
\vspace{-1ex}
\section{Related Work}

\noindent
\textbf{LLM benchmarking.}
Existing domain-specific LLM benchmarks fall into two categories. The first 
evaluates knowledge-based Q\&A~\cite{lu2024mathvista, singhal2023large, osvbench, hendrycks2021measuring, prakash2025quarch} but does not assess end-to-end engineering tasks.
The second 
evaluates end-to-end tasks~\cite{jimenez2023swe, thakur2023verigen,chen2021codex, roziere2023code, rtllm, mathai2024kgym}. 
None of them systematically cover \textit{fs} development tasks.
Moreover, they focused on providing a representative set of tasks, without deliberately designing and organizing tasks for analyzing model capabilities and limitations.
\pname{} is the first LLM benchmark for \textit{fs} development, enabling us to conduct a comprehensive study to break down the capabilities and limitations of LLMs.
\noindent
\textbf{LLM for OS development.}
Recent works have shown great promise for using LLMs to automate OS development~\cite{specfs:fast25, linux_llm_bug_fix, mathai2024kgym, osvbench,lau2026zerodaybenchevaluatingllmagents}.
AutoOS~\cite{autoos24} utilizes LLMs to tune kernel parameters.
SYSSPEC~\cite{specfs:fast25} allows developers to write formal specifications to guide LLM to generate a file system, but it still requires intensive manual prompting. 
This highlights the critical need to systematically analyze LLMs' ability to autonomously design and implement low-level software systems. Our study evaluates LLMs on \textit{fs} development, facilitating future research on AI-driven OS development.

\noindent
\textbf{File system design and implementation.}
File system development has long been notoriously labor-intensive, often requiring months to years of engineering effort~\cite{specfs:fast25, lee2015f2fs, ArpaciDusseau23-Book, bhat2017scaling}. File systems have been continuously evolving to exploit new hardware characteristics, such as the shift from FFS for HDDs to F2FS for flash-based SSDs~\cite{lee2015f2fs, ArpaciDusseau23-Book}.
Recent advancements in LLMs pave the way for AI-driven \textit{fs} development, yet it is still unclear how LLMs should be applied in this domain.
\pname{} introduces the first benchmark and conducts an empirical study to analyze LLM capabilities on \textit{fs} development.

\section{Conclusion}
\label{sec:conclusion}

We present \pname{}, an LLM benchmarking framework for file system design and implementation tasks.
With \pname{}, we evaluate various representative LLMs. We wish our study results would shed light on future research and development of using LLMs for file system development.

\bibliographystyle{ACM-Reference-Format}
\bibliography{ref}


\end{document}